\documentclass[usenatbib]{mn2e}
\usepackage{graphicx}

\title[SuperWASP-N Extra-solar Planet Candidates]
{SuperWASP-N Extra-solar Planet Candidates from Fields 06hr $<$ RA $<$ 16hr}
\author[S.R. Kane et al.]{
S.R. Kane$^{1,2}$,
W.I. Clarkson$^{4,9}$,
R.G. West$^{8}$,
D.M. Wilson$^{5}$,
D.J. Christian$^{3}$,
\newauthor
A. Collier Cameron$^{1}$,
B. Enoch$^{4}$,
T.A. Lister$^{1,5,11}$,
R.A. Street$^{3,11}$,
A. Evans$^{5}$,
\newauthor
A. Fitzsimmons$^{3}$,
C.A. Haswell$^{4}$,
C. Hellier$^{5}$,
S.T. Hodgkin$^{6}$,
K. Horne$^{1}$,
\newauthor
J. Irwin$^{6}$,
F.P. Keenan$^{3}$,
A.J. Norton$^{4}$,
J. Osborne$^{8}$,
N.R. Parley$^{4}$,
\newauthor
D.L. Pollacco$^{3}$,
R. Ryans$^{3}$,
I. Skillen$^{7}$,
P.J. Wheatley$^{10}$\\
$^1$School of Physics \& Astronomy, University of St Andrews, North Haugh,
St Andrews, Fife KY16 9SS, UK\\
$^2$Department of Astronomy, University of Florida, 211 Bryant Space Science
Center, Gainesville, FL 32611-2055, USA\\
$^3$School of Mathematics and Physics, Queen's University, Belfast,
University Road, Belfast, BT7 1NN, UK\\
$^4$Department of Physics \& Astronomy, The Open University, Milton Keynes
MK7 6AA, UK\\
$^5$Astrophysics Group, School of Chemistry \& Physics, Keele University,
Staffordshire ST5 5BG, UK\\
$^6$Institute of Astronomy, University of Cambridge, Madingley Road, Cambridge
CB3 0HA, UK\\
$^7$Isaac Newton Group of Telescopes, Apartado de correos 321, E-38700 Santa
Cruz de la Palma, Tenerife, Spain\\
$^8$Department of Physics \& Astronomy, University of Leicester, Leicester
LE1 7RH, UK\\
$^9$Space Telescope Science Institute, 3700 San Martin Drive, Baltimore, MD
21218, USA\\
$^{10}$Department of Physics, University of Warwick, Coventry CV4 7AL, UK\\
$^{11}$Las Cumbres Observatory Global Telescope, Goleta, CA 93117, USA
}

\begin{document}

\maketitle

\begin{abstract}

The Wide Angle Search for Planets (WASP) survey currently operates two
installations, designated SuperWASP-N and SuperWASP-S, located in the
northern and southern hemispheres respectively. These installations are
designed to provide high time-resolution photometry for the purpose of
detecting transiting extra-solar planets, asteroids, and transient events.
Here we present results from a transit-hunting observing campaign using
SuperWASP-N covering a right ascension range of 06hr $<$ RA $<$ 16hr.
This paper represents the fifth and final in the series of transit
candidates released from the 2004 observing season. In total, 729,335
stars from 33 fields were monitored with 130,566 having sufficient
precision to be scanned for transit signatures. Using a robust transit
detection algorithm and selection criteria, 6 stars were found to have
events consistent with the signature of a transiting extra-solar planet
based upon the photometry, including the known transiting planet XO-1b.
These transit candidates are presented here along with discussion of
follow-up observations and the expected number of candidates in relation
to the overall observing strategy.

\end{abstract}

\begin{keywords}
methods: data analysis -- planetary systems -- stars: variables: other
\end{keywords}

\section{Introduction}

More than 30 extra-solar planets are now known to transit their parent stars,
most of which were discovered through photometric monitoring. Transit surveys
which use a narrow-field and a large magnitude depth have had some success;
most notably that of the OGLE transiting planets (e.g., \citet{kon03}).
Surveys which utilise the ``wide and shallow'' technique of monitoring
relatively bright field stars have achieved recent rapid success, including
the discoveries of TrES-3 \citep{odo07}, XO-2b \citep{bur07}, and several
new HATNet planets (\citet{kov07} for example). The transit technique has
matured through overcoming serious obstacles which were impeding the data
analysis, such as improved optimal photometric methods for wide-field
detectors \citep{har04} and reduction of correlated (red) noise
\citep*{pon06,tam05}.

The Wide Angle Search for Planets (WASP) project currently operates two
SuperWASP instruments \citep{pol06}, one in the northern hemisphere on La
Palma (SuperWASP-N) and the other (SuperWASP-S) located at South African
Astronomical Observatory (SAAO). The relatively large field-of-view (FOV)
of the SuperWASP design allows each instrument to monitor a substantial
amount of the visible sky with relatively high time resolution. SuperWASP-N
has been acquiring data since late-2003, achieving photometric precision of
1\% for stars brighter than $V \sim 11.5$.

The process of extracting reliable transit signatures from the SuperWASP
data involves a systematic approach of removing trends from the photometry,
employing an efficient transit detection algorithm \citep{col06}, and
performing spectroscopic follow-up using both low-medium resolution and high
resolution spectrographs to identify false-positives and confirm planetary
candidates. This method has already been used successfully on SuperWASP data
to detect the planets WASP-1b and WASP-2b \citep{col07a}. However, this
process typically begins with millions of lightcurves from which many
transit candidates are found and provides an extremely useful reference
source for future surveys which monitor the same fields. For this reason,
the transit candidates detected in several RA ranges have been published for
the benefit of the transit survey community \citep{chr06,cla07,lis07,str07}.

We present the results of a photometric search for exoplanetary transits
using data from SuperWASP-N, covering the fields in the range 06hr $<$ RA $<$
16hr. Section 2 describes the 2003/2004 SuperWASP-N observing campaign, the
predicted transit recovery rate, and the data reduction techniques. Section 3
discusses the transit detection algorithm and the candidate selection
criteria. The results of this search are presented in section 4 with
discussion of the candidates which have been followed up and those which are
considered high priority. Section 5 discusses the results from the analysis
and the number of final candidates in relation to the limitations of the
survey for this particular RA range.

\section{Data Acquisition}

\subsection{Observations and Recovery Rate}

The SuperWASP-N instrument is a robotic observatory designed to provide
precision photometry for large areas of sky. First light was achieved in
November 2003 and observations have continued until the present time. The
instrument consists of a fork mount which is able to support up to eight
lens/detector combinations simultaneously, described in more detail in
\citet{pol06}. When the instrument is operating at full capacity, the sky
coverage becomes substantial leading to a total FOV of 482 square degrees.
with a pixel scale of 13.7 arcsec per pixel. During the 2003/2004
(hereafter referred to as 2004) observing season, five out of the eight
detectors were installed. Even so, the data rate from those five cameras
exceeded the electronic transfer capability of the ethernet connection on
La Palma at that time and so data was stored via a tape (DLT) autoloader
then mailed to data reduction sites.

The observing strategy during the 2004 observations was given careful
consideration, mostly in an effort to reduce the false-alarm rate due to
blended eclipsing binaries \citep{bro03}. This is a particularly important
problem to solve for instruments such as SuperWASP whose wide fields can
encompass a significant gradient in stellar densities and whose large pixel
sizes greatly increase the chance of blending. The fields monitored were
selected at 1 hour increments in RA and lying along $\mathrm{Dec} = +28\degr$.
This avoided the Galactic plane and hence over-crowding in the fields. In
addition, the ecliptic was avoided to reduce sky contamination from the Moon
and other bright Solar System objects. An exposure time of 30 seconds was
used for maximum dynamic range and observations cycled between up to 8
fields. The slew-time of the mount ensured that an observing cadence $\sim
8$ minutes per field was achieved. The resulting dataset contains
lightcurves for over 6.7 million stars in the magnitude range $8 < V < 15$.
These data were divided by RA range into six separate datasets which each
contained $\sim 1$ million stars.

\begin{table}
  \caption{Fields observed using SuperWASP-N during 2004 in the range 06hr
    $<$ RA $<$ 16hr. The table shows the total number of stars monitored per
    field, as well as the number of extracted stars for transit hunting $N_e$,
    the initial number of candidates identified by the detection algorithm
    $N_i$, and the final number of candidates $N_f$.}
  \begin{tabular}{@{}cccccccc}
    \hline
    RA & Dec & Nights & Frames & Stars & $N_e$ & $N_i$ & $N_f$ \\
    \hline
    0616 & +3126 &  37 &  627 & 128284 & 23936 & 280 & 0 \\
    1043 & +3126 &  31 &  653 &  15014 &  2775 &  61 & 0 \\
    1044 & +2427 &  32 &  654 &  16365 &  2785 &  18 & 0 \\
    1116 & +3126 &  31 &  584 &  14234 &  2459 &  72 & 0 \\
    1117 & +2326 &  31 &  653 &  15521 &  2586 &  39 & 0 \\
    1143 & +3126 &  51 & 1200 &  13907 &  2508 &  80 & 0 \\
    1144 & +2427 &  52 & 1203 &  15400 &  2591 &  78 & 1 \\
    1144 & +3944 &  45 &  908 &  14985 &  2285 &  32 & 0 \\
    1216 & +3126 &  51 & 1111 &  15177 &  2592 &  77 & 0 \\
    1217 & +2326 &  51 & 1200 &  13045 &  2460 &  82 & 0 \\
    1243 & +3126 &  86 & 2378 &  15082 &  2605 & 115 & 1 \\
    1244 & +2427 &  85 & 2382 &  13287 &  2547 & 161 & 0 \\
    1244 & +3944 &  78 & 1985 &  15065 &  2410 & 111 & 0 \\
    1316 & +3126 &  86 & 2277 &  15259 &  2566 & 147 & 0 \\
    1317 & +2326 &  87 & 2386 &  12207 &  2408 & 134 & 0 \\
    1342 & +3824 & 103 & 2846 &  14792 &  2724 & 139 & 0 \\
    1342 & +4642 & 101 & 2689 &  15928 &  2556 & 155 & 0 \\
    1343 & +3126 & 101 & 2842 &  15357 &  2616 & 115 & 0 \\
    1417 & +3824 & 102 & 2848 &  15509 &  2871 & 154 & 0 \\
    1418 & +3025 & 103 & 2854 &  16374 &  2933 & 167 & 0 \\
    1443 & +3126 & 125 & 4018 &  16847 &  3071 & 167 & 0 \\
    1444 & +2427 & 113 & 3918 &  19502 &  3180 & 131 & 0 \\
    1444 & +3944 & 117 & 3503 &  17476 &  3285 & 289 & 0 \\
    1516 & +3126 & 123 & 3881 &  20125 &  3448 & 253 & 0 \\
    1517 & +2326 & 123 & 3986 &  21242 &  3468 & 245 & 1 \\
    1543 & +3126 & 130 & 4775 &  23180 &  3963 & 208 & 0 \\
    1544 & +2427 & 113 & 4560 &  22737 &  3907 & 179 & 0 \\
    1544 & +3944 & 122 & 4271 &  20097 &  3874 & 296 & 0 \\
    1616 & +3126 & 127 & 4626 &  28627 &  5133 & 381 & 2 \\
    1617 & +2326 & 129 & 4722 &  30043 &  5110 & 433 & 1 \\
    1643 & +3126 & 129 & 5333 &  35443 &  6233 & 363 & 0 \\
    1644 & +2427 & 112 & 4969 &  33266 &  6378 & 274 & 0 \\
    1644 & +3944 & 121 & 4883 &  29958 &  6303 &   9 & 0 \\
    \hline
  \end{tabular}
\end{table}

This paper presents the analysis of the stars in the range
06hr~$<$~RA~$<$~16hr. A total of 1,832,777 stars were monitored in this
range over 56 fields and 95,526 frames were acquired. For each field, a
series of constraints were applied to extract suitable stars for
transit hunting. These include extracting only stars whose lightcurves have
an RMS precision better than 0.01 mag and whose baseline contains at least
500 frames spread over at least 10 nights. A summary of the fields observed
in this dataset is shown in Table 1. Of the 56 fields, 23 did not meet the
baseline criteria for lightcurve extraction and so are not included in
Table 1. A total of 130,566 stars were extracted for transit hunting from
the 729,335 remaining stars using the above criteria.

The lack of baseline coverage suffered by many of the fields due to the
visibility of the fields over the observing campaign was particularly acute
at small RA. Consequently, the sensitivity to planetary transits varies
greatly over the RA range in this dataset. This is illustrated by the transit
recovery plots shown in Figure 1. Transit signatures were randomly generated
with periods in the range $1 < P < 5$ and injected into the times of
observation for each field. A transit signature is considered ``recovered''
if both the ingress and egress of the transit is observed. Figure 1 shows the
results of this simulation for four fields which span the range of RA in this
dataset; in each case using 2, 4, and 6 transits as the requirement for
detection. The reduction in transit recovery rate is clearly quite dramatic
for field which are observed for less than $\sim 60$ nights.

\begin{figure*}
  \begin{center}
    \begin{tabular}{cc}
      \includegraphics[angle=270,width=8.2cm]{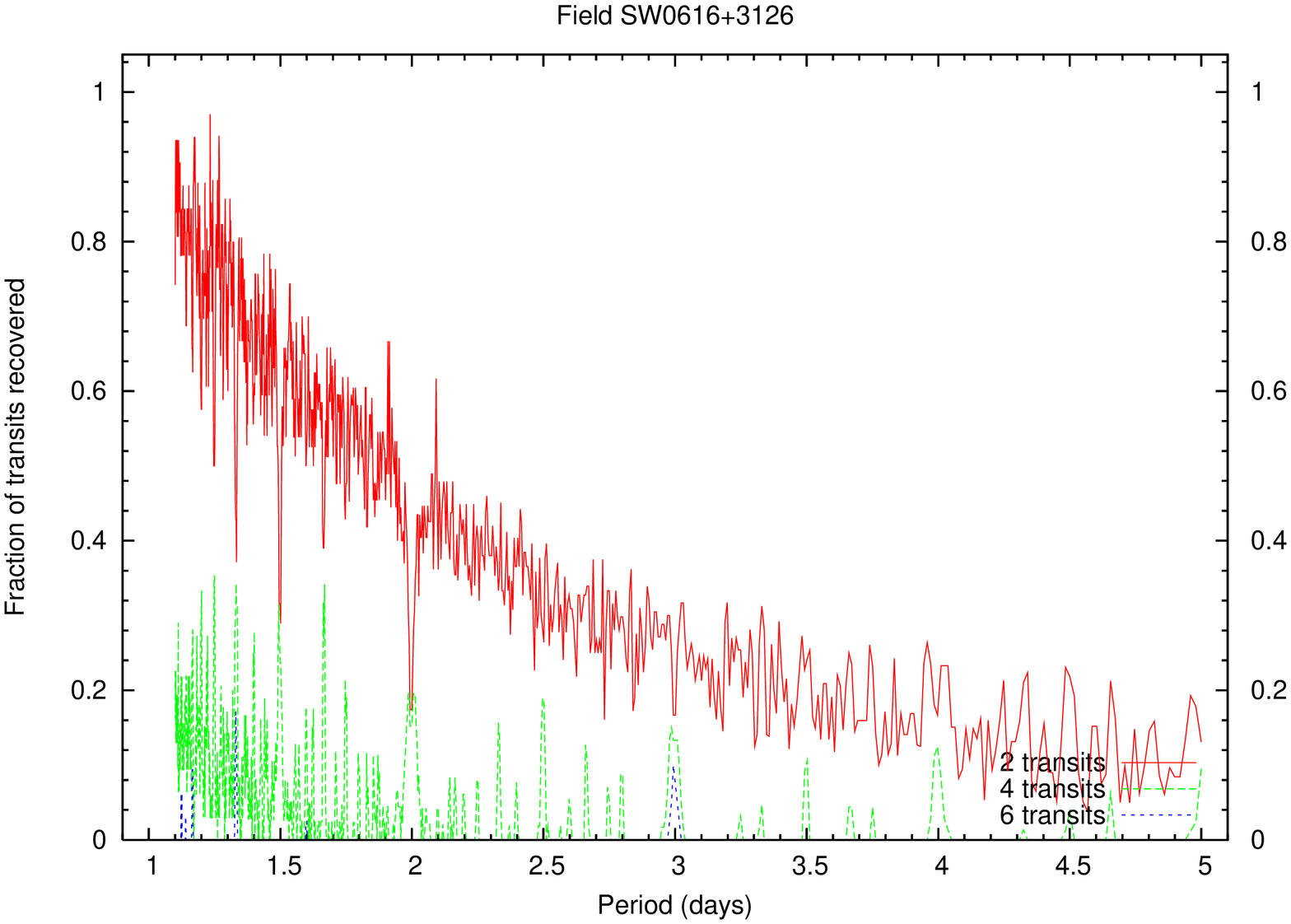} &
      \includegraphics[angle=270,width=8.2cm]{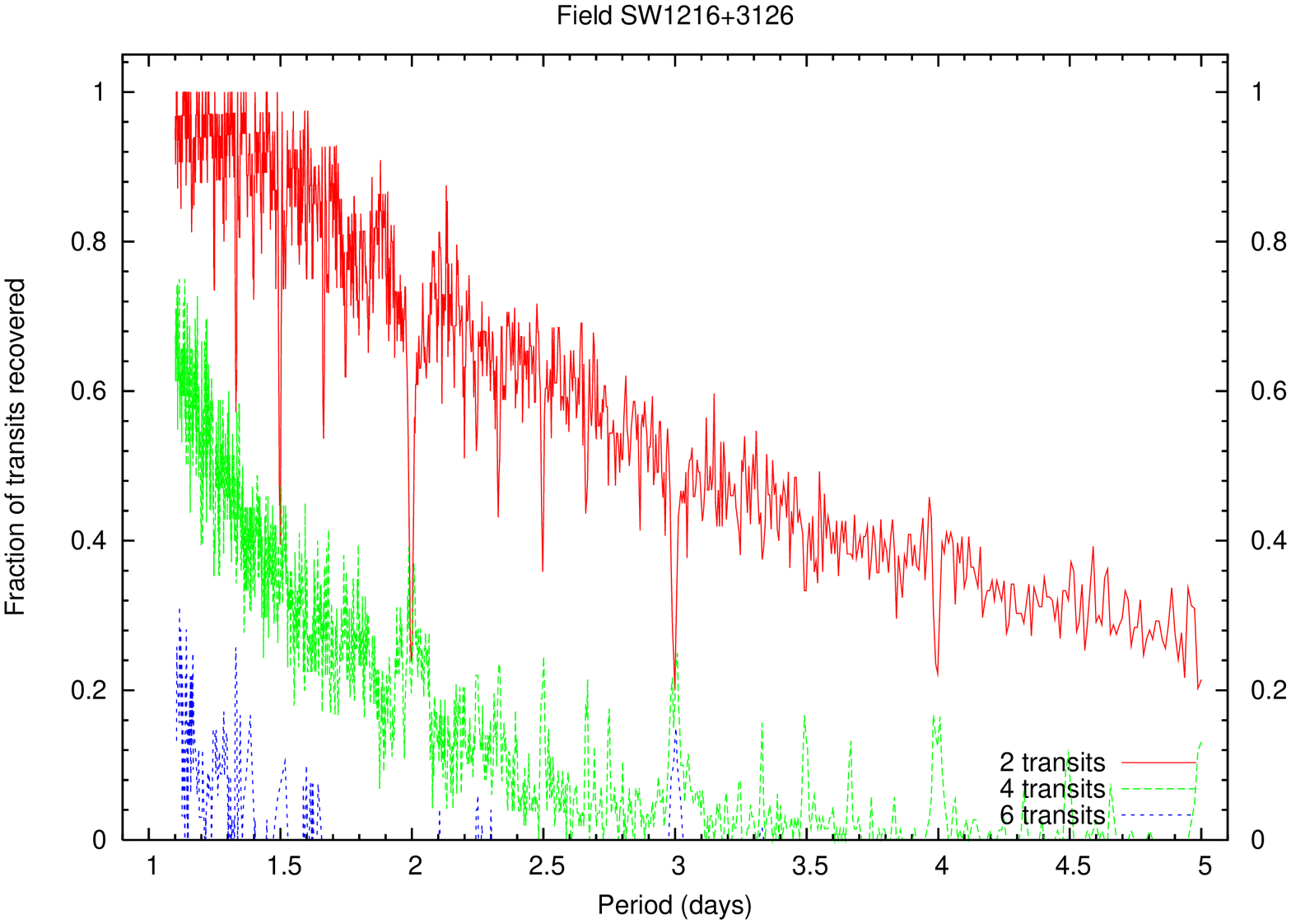} \\
      \includegraphics[angle=270,width=8.2cm]{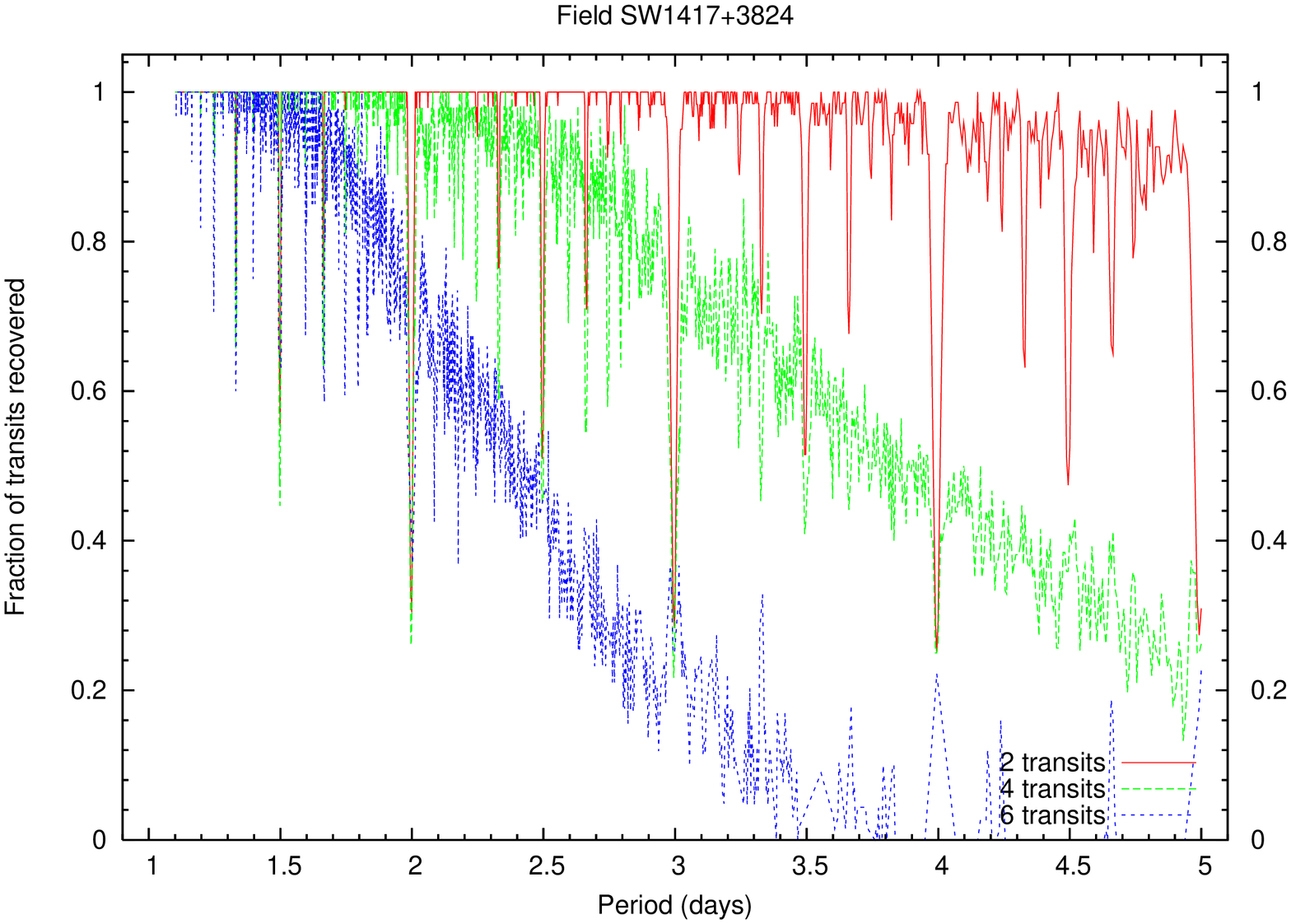} &
      \includegraphics[angle=270,width=8.2cm]{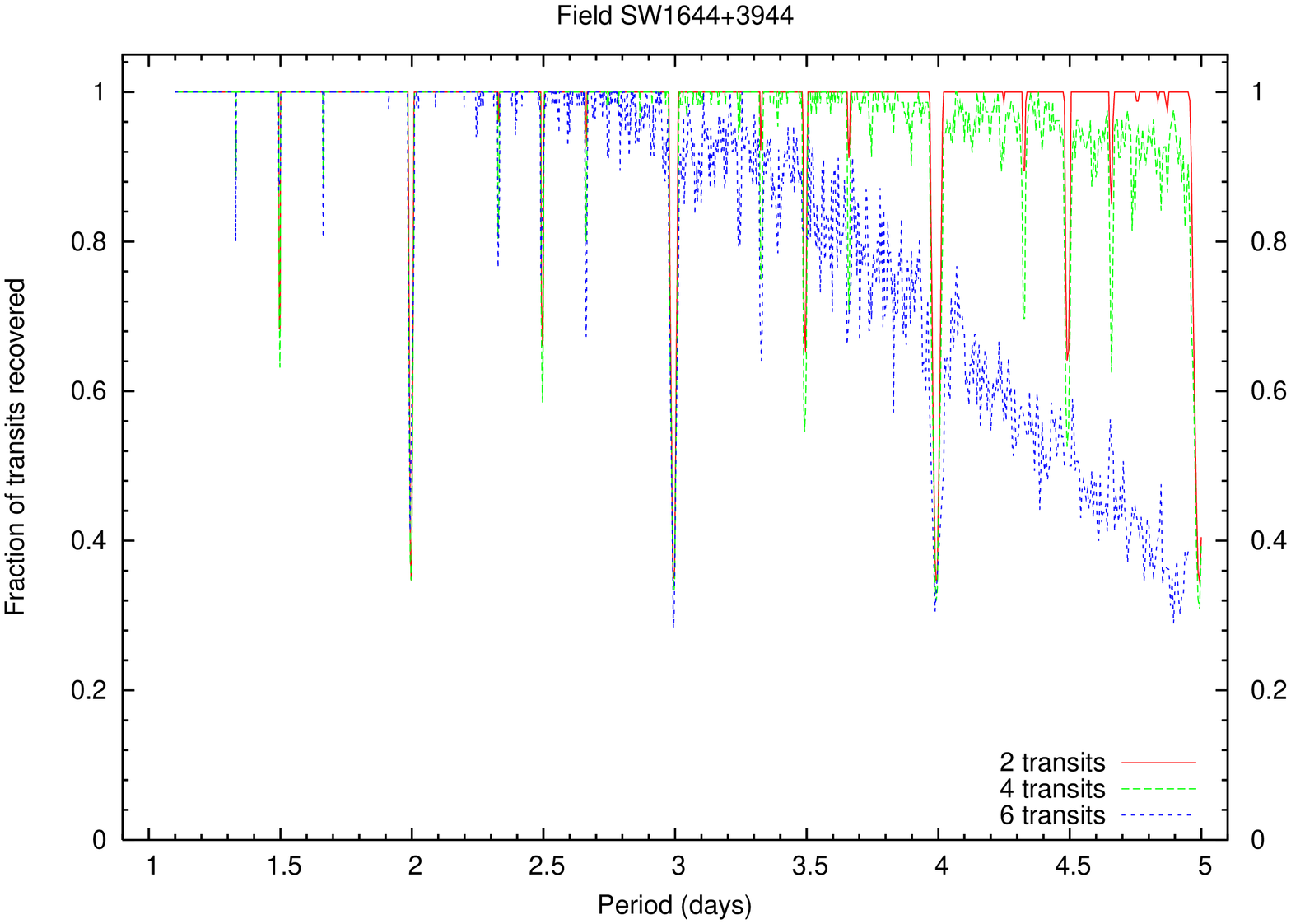} \\
    \end{tabular}
  \end{center}
  \caption{Transit recovery rate for the fields SW0616+3126, SW1216+3126,
    SW1417+3824, and SW1644+3944; observed for 37, 51, 102, and 121 nights
    respectively. The solid line represents the recovery of 2 transits, the
    dashed line 4 transits, and the dotted line 6 transits.}
\end{figure*}

\subsection{Data Reduction}

Reduction of the SuperWASP-N data required producing milli-magnitude
photometry whilst managing a high data rate. An automated reduction
pipeline was constructed which was able to achieve this goal and is
described in more detail in \citet{pol06}. This description will
concentrate on the main factors to consider when reducing wide-field data
from instruments such as SuperWASP. Wide-field issues such as vignetting,
point-spread function (PSF) distortion, and spatial dependence were
encountered with the WASP0 prototype instrument \citep{kan04}. The WASP0
dataset provided an excellent starting point around which to solve the
same wide-field issues that would affect the SuperWASP data and pipeline.

The calibration frames include bias, dark, and flat-field frames which
are generally acquired on a nightly basis when the automated enclosure
commences operations at dusk. These frames are handled by the pipeline
through a series of statistical tests which are able to classify them
and create master frames. To create the master flat-field, a vignetting
map and a shutter time correction map are created and then combined with
the flat-field through an inverse variance weighted linear least-squares
fit. Iterative sigma-clipping and smoothing via spline-fitting leads to
an accurate representation of the sky brightness for the wide-field.

Rather than fit the spatially-variable PSF shape of the stellar images,
weighted aperture photometry is used to compute the flux in a circular
aperture of tunable radius. This is achieved by implementing a
flux-weighted astrometric fit which uses an automatically extracted
subset of the Tycho-2 \citep{hog00} catalogue, then fixing the aperture
locations based on entries within the USNO-B1.0 \citep{mon03} catalogue
with a 2nd epoch red magnitude brighter than 15.0. This also allows the
correction of spatially-dependent aspects which are normally assumed to
be constant across the frame, for example the airmass and heliocentric
time correction. The vignetting and barrel distortion produced by the
camera optics can result in serious blending effects for stars whose
neighbours possess significantly distorted stellar profiles. It has been
shown by \citet{bro03} and \citet{tor04} that blending can have a
detrimental effect on transit searches. Stars which are significantly
affected by blending are identified by computing blending indices
$B_1 = (F_3 - F_1) / F_1$ and $B_2 = (F_3 - F_2) / F_2$ where $F_1$,
$F_2$, and $F_3$ are the flux measurements from aperture radii of 2.5,
3.5, and 4.5 respectively. The comparison of these indices leads to an
effective exclusion of blended stars as described by \citet{kan05}.

The photometric data which is produced by the pipeline is further refined
by applying corrections for primary and secondary extinction through an
iterative process which is explained in detail by \citet{col06}. The
instrumental magnitudes are then transformed to Tycho-2 magnitudes by
using local calibrators observed on exceptional nights. These corrected
data are stored as FITS binary tables and ingested into the SuperWASP
Data Archive which is located at the University of Leicester.

\section{Hunting for Transits}

This section describes the methodology used to sift the archived
photometric data for transit signatures. The description here is based
upon the principles suggested by \citet{col06} and \citet{col07b} with
applications to these particular data in mind.

\subsection{De-trending Lightcurves}

Achieving the photometric accuracy necessary to be sensitive to transiting
extra-solar planets is a major challenge for wide-field instruments such
as SuperWASP. The stellar lightcurves which are extracted from the archive
typically have systematic errors which have the strong potential to produce
false-alarms when scanned for transit signatures. The severity of the
effects of correlated noise on the planet yield was demonstrated by
\citet{pon06}. The SysRem algorithm proposed by \citet{tam05} is effective
at identifying and removing correlated noise and was the method adopted
for the SuperWASP data prior to transit hunting.

As shown by \citet{col06}, the SysRem algorithm identified four basis
functions, each of which represents a distinct systematic noise pattern,
which were used to model the global systematic errors. The frequency
structure of the noise outside of transit was found to be correlated and
characterised by a power-law. As a result of this frequency dependence,
the entire correlated noise is referred to as red noise. Applying the
deduced correction to the SuperWASP data reduced the RMS amplitude of the
red noise from 0.0025 magnitude to 0.0015 magnitudes. These basis
functions were representations of such effects as colour corrections,
extinction, and the vignetting function. Further basis functions were
identified but these were not applied to the error model to avoid the
risk of removing real variability. The resulting de-trended lightcurves
are then ready to be analysed for periodic variability including transit
signatures.

\subsection{Transit Detection Algorithm}

The automation of transit detection algorithms for very large datasets has
presented a major challenge for the transit survey teams, prompting
extensive studies into optimal solutions (eg., \citet*{def01,kov02}). The
transit model which is applied needs to be optimised to correctly balance
the computational time and the efficiency of avoiding false-positive
detections. The hybrid algorithm suggested by \citet{col06} uses a refined
version of the Box Least-Squares (BLS) algorithm to perform a search on a
coarse grid of transit epochs. The algorithm then rejects candidates based
upon such features as strong variability, the number of transits, and
significant gaps in the phase-folded lightcurve. This removes more than
95\% of the search sample, allowing for a finer grid search on the
remaining stars using the Newton-Raphson method of \citet*{pro05}. Recall
also from section 2.1 that a requirement imposed before transit searching
was that each star have $\geq 500$ measurements obtained over $\geq 10$
nights with an RMS precision $\leq 0.01$ mag. In practise, the RMS
restriction results in the inclusion of stars with Tycho V $\la 13$.

For the SuperWASP fields, a coarse grid search was conducted with periods
in the range $0.9 \leq P \leq 5.0$. The fine grid search to the remaining
stars yields a candidate list which includes measurements of basic transit
fit parameters; such as the period, duration, depth, and epoch of first
transit. Additionally provided are the delta chi-square of the fit $\Delta
\chi^2$, the ratio of the best-fit transit model to that of the best-fit
anti-transit model $\Delta \chi^2 / \Delta \chi_{-}^2$ \citep{bur06}, the
signal to red noise ratio $S_{red}$ (as described by \citet{pon06}), and
the signal-to-noise of the ellipsoidal variation $\mathrm{S/N}_{ellip}$
to reduce false-alarms due to eclipsing binaries \citep{sir03}. The
combination of these statistics allows for a powerful analysis tool in
sifting transit candidates from the list.

\subsection{Candidate Selection Criteria}

The evolution of the candidate list provided by the detection algorithm
and the final list of candidates proceeds in three distinct stages. The
first stage is the visual inspection of the folded lightcurves, a process
by which we are able to discard many of the candidates for which there is
clear evidence of a secondary eclipse. In addition, candidates whose
best-fit period is closely correlated with an integer number of days are
rejected in order to minimise the number of aliases. This stage is also
used to discard those whose data are of exceptionally poor quality. It
should be emphasised that this step is not to select transit candidates,
rather it is to reject obvious false-alarms. Through this visual
inspection stage, also employed using identical criteria by the other
SuperWASP-N 2004 candidate releases \citep{chr06,cla07,lis07,str07},
generally over 95\% of the candidates are rejected.

The second stage of candidate sifting uses the quantities computed by the
detection algorithm. Candidates were systematically removed from the list
if any of the following are true: (i) the signal to red noise ratio is low
($S_{red} < 8.0$), (ii) there are significant $\sim 1$ days aliases ($P <
1.05$ days), (iii) less than 3 transits are observed, (iv) the anti-transit
ratio is low ($\Delta \chi^2 / \Delta \chi_{-}^2 < 2.0$), or (v) the
ellipsoidal variation is high ($\mathrm{S/N}_{ellip} > 8.0$). The surviving
candidates are combined together and sorted by RA so that any close
companions can be easily identified.

The third stage ingests the candidates from the second stage into the
Variable Star Investigator (VSI) which is an automated query tool
developed by the SuperWASP consortium to provide colour information from
existing photometric catalogues. The VSI tool returns photometric
information from such catalogues as Tycho-2 and 2MASS \citep{skr06} via
VIZIER \citep*{och00}. The information is used to provide estimates of
basic stellar parameters, such as spectral type, effective temperature
(see \citet{str07}), and radius $R_\star$. The transit depth then
translates into a direct estimate of the planetary radius $R_p$. The
transit parameters allow a calculation of the exoplanet diagnostic
parameter $\eta_p$ which serves as a measure of the candidate reliability
and is described by \citet{tin05}. Moreover, findercharts from DSS
\citep{cab03} are used to reveal the presence of any close companions
within a 48\arcsec aperture. Candidates are removed from the list if
either (i) there is a brighter object within the 48\arcsec aperture, or
(ii) if the estimated radius is too large ($R_p \ga 1.6 R_J$). This
radius cutoff was selected to remain consistent with the previous
SuperWASP candidate papers and also the relatively large radius of the
recently discovered planet TrES-4 \citep{man07}.

\section{Results}

This section presents results from the transit search for the 33 fields
monitored in the 6hr $<$ RA $<$ 16hr range which met the search criteria,
including the candidate rejection process and description of the
surviving candidates.

\subsection{Planetary Transit Search}

As described in section 2.1, we required that a number of conditions be
satisfied before subjecting the data to the transit search algorithm. A
total of 729,335 stars from 33 fields achieved the baseline requirement
for transit hunting, and 130,566 of these stars also met the photometric
precision requirement. These stars were processed by the detection
algorithm, yielding an output of fit parameters and lightcurves folded
on the best-fit period. This output contained 5,445 stars which were
selected by the detection algorithm as having transit signatures that
require further investigation. Table 1 lists the number of stars
extracted for transit hunting $N_e$, the initial number of candidates
detected $N_i$, and the final number of candidates $N_f$ per field.

The first stage (visual inspection) of the candidate sifting process
yielded 36 candidates from the original list of 5,445. Table 2 presents
these candidates in order of increasing RA. Only those candidates for
which a clear transit was visible were selected to populate this list.
The criteria described in section 3.3 for the second stage of candidate
sifting were then applied to this list, using the information displayed
in the table. The code in the final column of the table shows which test
from the second stage was failed by that star. In the instance where
more than one test is failed by a star, the first test failed is shown.
Half of those which failed to pass through the second stage were
eliminated due to the degree of ellipsoidal variation in the
out-of-transit lightcurve, emphasising the potential contamination by
grazing eclipsing binary stars.

\begin{table*}
  \caption{List of candidates that passed the visual inspection (first stage)
    of the transit candidates detected by the BLS algorithm. The final column
    lists a rejection code for those which were rejected according to the
    information available from the BLS search, as described in section 3.3.
    The codes are as follow: (R) signal to red noise ratio $S_{red}$, (P)
    period, (A) anti-transit ratio, and (E) signal-to-noise of ellipsoidal
    variation $\mathrm{S/N}_{ellip}$.}
  \begin{tabular}{@{}ccccccccccc}
    \hline
    SWASP ID    & Period & Depth & Duration & Epoch        & $N_t$ &
    $\Delta \chi^2$ & $\Delta \chi^2 / \Delta \chi_{-}^2$ &
    $\mathrm{S/N}_{ellip}$ & $S_{red}$ & Code \\
    (1SWASP$+$) & (days) & (mag) & (hours)  & (2450000$+$) &       &
                    &                                     &
                           &           &      \\
    \hline
J113655.81+281708.5 & 0.960430 & 0.0344 & 2.592 & 3127.5413 &  8 &   336.2065 &  5.3844 &  3.953 &  9.181 & P \\
J115418.56+351211.5 & 1.427904 & 0.2355 & 2.184 & 3127.9543 &  6 &  1672.4440 & 14.0745 &  3.814 & 15.175 &  \\
J115718.66+261906.1 & 1.226804 & 0.0170 & 2.784 & 3128.1643 &  7 &   140.3018 &  7.6224 &  2.214 & 12.916 &  \\
J120214.02+332920.4 & 2.589406 & 0.1456 & 3.696 & 3126.5193 &  5 &  3569.5464 & 12.8868 &  2.515 & 10.962 &  \\
J120933.96+335637.2 & 1.167674 & 0.0278 & 2.040 & 3128.1484 &  8 &  1131.8057 &  1.5813 & 35.972 & 10.319 & A \\
J121653.45+263804.1 & 0.982613 & 0.0301 & 1.152 & 3128.3462 &  3 &    90.3980 &  5.2886 &  1.985 & 12.937 & P \\
J122557.80+334651.1 & 1.362462 & 0.0811 & 2.568 & 3127.7017 & 13 &  4330.1172 & 24.3831 & 15.725 & 14.670 & E \\
J122600.12+274221.5 & 1.366517 & 0.0265 & 1.896 & 3127.6277 &  5 &   124.4934 &  2.7044 &  3.387 &  7.215 & R \\
J122640.99+310203.0 & 1.337858 & 0.0139 & 1.368 & 3127.8386 &  7 &    63.7055 &  1.8859 &  0.133 & 10.476 & A \\
J122812.01+324132.3 & 3.090437 & 0.0874 & 2.376 & 3126.7476 &  4 &  2586.7490 & 24.3444 & 11.980 & 20.691 & E \\
J124428.08+255631.5 & 2.763064 & 0.0547 & 1.800 & 3127.8606 &  4 &    80.7336 &  1.6915 &  0.031 & 19.010 & A \\
J125704.45+320657.4 & 2.899515 & 0.0196 & 2.688 & 3127.5493 &  6 &    36.5989 &  1.3138 &  1.670 &  8.645 & A \\
J130303.23+423223.7 & 2.858061 & 0.0829 & 2.208 & 3125.7209 &  5 &  1505.2094 &  3.2609 &  5.610 & 10.246 &  \\
J130322.00+350525.4 & 2.674207 & 0.0179 & 3.264 & 3127.3608 &  6 &   304.3417 &  5.6491 &  2.001 &  8.467 &  \\
J130409.52+201138.5 & 1.084250 & 0.1161 & 1.200 & 3127.5449 & 10 &  1827.6654 & 12.9069 &  8.689 & 13.746 & E \\
J133022.79+330746.7 & 3.626903 & 0.0863 & 2.568 & 3125.3120 &  5 &  1229.3250 & 35.5255 &  0.677 & 10.713 &  \\
J133156.81+460026.6 & 3.166482 & 0.0844 & 5.088 & 3128.2366 & 10 &  1924.8636 & 10.2625 &  5.186 & 12.102 &  \\
J133623.52+283745.3 & 0.959561 & 0.0586 & 2.712 & 3127.6365 & 25 &  3638.7996 &  8.4634 & 22.666 & 15.258 & P \\
J141558.71+400026.7 & 2.450924 & 0.0397 & 1.632 & 3126.7324 &  8 &   281.4857 &  3.5665 &  2.999 & 13.474 &  \\
J142947.03+230708.4 & 4.219485 & 0.0868 & 2.976 & 3125.0061 &  8 &  2755.3987 & 32.1253 &  7.453 & 18.133 &  \\
J144659.77+285248.3 & 3.798768 & 0.1094 & 3.048 & 3125.3835 & 11 &  3284.5603 & 27.9419 &  4.250 & 20.591 &  \\
J151508.36+301413.7 & 1.394774 & 0.2315 & 2.280 & 3128.1123 & 15 & 13151.5469 & 11.0242 &  8.328 & 23.730 & E \\
J152131.01+213521.3 & 1.338018 & 0.0259 & 2.952 & 3127.7908 & 21 &   579.2580 &  7.1841 &  1.342 & 13.865 &  \\
J152645.62+310204.3 & 1.409102 & 0.0297 & 1.248 & 3127.3904 & 16 &  1648.0901 &  3.5165 & 13.729 & 12.944 & E \\
J153135.51+305957.1 & 4.467224 & 0.0367 & 4.080 & 3128.3252 & 10 &  1652.4308 & 17.8900 &  5.468 & 18.038 &  \\
J153741.83+344433.4 & 0.963514 & 0.0296 & 1.320 & 3127.8230 & 24 &   253.6837 &  2.0412 &  2.376 & 19.644 & P \\
J160211.83+281010.4 & 3.941554 & 0.0238 & 2.352 & 3127.0391 & 11 &  1176.7375 &  3.6554 &  1.896 & 13.653 &  \\
J160242.43+290850.1 & 1.304693 & 0.0454 & 1.872 & 3127.2471 & 23 &  1878.1125 &  2.5227 &  4.079 & 14.767 &  \\
J160944.95+202609.7 & 1.644243 & 0.0917 & 2.784 & 3127.7603 & 16 & 20028.2793 & 25.3475 &  8.789 & 15.681 & E \\
J161644.68+200806.8 & 3.967135 & 0.1608 & 4.176 & 3124.4480 &  8 &  3263.7249 & 32.0755 &  5.043 & 30.317 &  \\
J161732.90+242119.0 & 1.453738 & 0.0157 & 1.440 & 3127.6799 & 16 &   325.0795 &  4.4974 &  0.801 & 12.968 &  \\
J162437.86+345723.8 & 4.423851 & 0.0238 & 2.952 & 3124.5613 &  6 &   514.2166 & 11.8262 &  0.755 & 19.494 &  \\
J163245.61+321754.9 & 3.538188 & 0.0928 & 2.808 & 3128.3401 & 10 & 24310.1016 & 72.4177 & 13.176 & 23.421 & E \\
J163844.53+411849.0 & 3.859897 & 0.1098 & 3.552 & 3125.8313 & 10 &  7159.1206 & 26.1351 & 10.467 & 39.587 & E \\
J165424.59+241318.7 & 2.571173 & 0.0434 & 2.160 & 3127.1179 & 14 &   755.8041 & 10.8147 &  2.247 & 16.440 &  \\
J165949.13+265346.1 & 2.682413 & 0.0206 & 1.848 & 3128.1721 & 11 &   957.7991 &  4.5615 & 14.616 & 14.128 & E \\
    \hline
  \end{tabular}
\end{table*}

The 18 candidates which passed through the second stage were then subjected
to further analysis using VSI. These candidates are listed in Table 3 along
with the relevant information extracted from VSI, including the number of
brighter stars $N_{bri}$ and fainter stars $N_{fai}$ ($< 5$ mag) within the
48\arcsec aperture. Since the Tycho-2 catalogue is incomplete below $V \sim
11.5$, the USNO-B1.0 catalogue is used to identify individual objects within
the aperture. Two of the candidates were rejected due to their sharing the
aperture with at least one brighter object. The colours from the photometric
catalogues returned by VSI combined with the fit parameters provided by the
detection algorithm are sufficient to calculate approximate values of stellar
parameters for the host star. The code column of Table 3 shows that most of
the candidates have a predicted planet size that is significantly larger than
one would expect, and are therefore excluded from the final list.

\begin{table*}
  \caption{List of candidates that passed the second stage tests and were
    subsequently subjected to catalogue-based tests, as described in section 3.3.
    The rejection codes shown in the final column are as follows: (B) brighter
    object within the specified aperture, and (S) the estimated size (radius) of
    the planet. A total of six candidates pass all of the tests.}
  \begin{tabular}{@{}ccccccccccc}
    \hline
    SWASP ID    & $V_{SW}$ & $V_{SW}-K$ & $J-H$ & $H-K$ & $R_\star$   & $R_p$   &
    $\eta_p$ & $N_{bri}$ & $N_{fai}$ & Code\\
    (1SWASP$+$) &          &            &       &       & ($R_\odot$) & ($R_J$) &
             &           &           &     \\
    \hline
J115418.56+351211.5 & 12.994 & 1.16 & 0.329 & 0.070 & 1.00 & 5.47 & 0.64 & 0 & 1 & S \\
J115718.66+261906.1 & 11.116 & 1.06 & 0.177 & 0.037 & 1.32 & 1.55 & 1.06 & 0 & 2 &  \\
J120214.02+332920.4 & 12.533 & 1.63 & 0.290 & 0.060 & 1.08 & 3.23 & 1.09 & 0 & 0 & S \\
J130303.23+423223.7 & 11.657 & 1.56 & 0.235 & 0.055 & 1.21 & 2.58 & 0.67 & 0 & 1 & S \\
J130322.00+350525.4 & 10.893 & 1.77 & 0.275 & 0.099 & 1.12 & 1.06 & 1.21 & 0 & 0 &  \\
J133022.79+330746.7 & 12.529 & 1.78 & 0.268 & 0.081 & 1.12 & 2.33 & 0.76 & 0 & 0 & S \\
J133156.81+460026.6 & 12.627 & 2.13 & 0.466 & 0.104 & 0.78 & 1.98 & 1.73 & 0 & 1 & S \\
J141558.71+400026.7 & 12.393 & 1.44 & 0.211 & 0.063 & 1.25 & 1.90 & 0.53 & 0 & 0 & S \\
J142947.03+230708.4 & 12.149 & 1.57 & 0.281 & 0.060 & 1.10 & 2.61 & 0.79 & 0 & 0 & S \\
J144659.77+285248.3 & 12.726 & 1.73 & 0.286 & 0.084 & 1.08 & 2.68 & 0.86 & 0 & 1 & S \\
J152131.01+213521.3 & 12.188 & 1.36 & 0.240 & 0.046 & 1.19 & 1.62 & 1.17 & 0 & 1 &  \\
J153135.51+305957.1 & 11.778 & 1.28 & 0.252 & 0.060 & 1.16 & 2.03 & 1.00 & 0 & 1 & S \\
J160211.83+281010.4 & 11.319 & 1.79 & 0.338 & 0.074 & 0.98 & 1.21 & 0.76 & 0 & 0 &  \\
J160242.43+290850.1 & 12.417 & 1.99 & 0.380 & 0.051 & 0.91 & 1.53 & 0.88 & 0 & 1 &  \\
J161644.68+200806.8 & 12.352 & 1.77 & 0.587 & 0.156 & 0.69 & 3.18 & 1.12 & 1 & 3 & B \\
J161732.90+242119.0 & 11.959 & 2.61 & 0.477 & 0.122 & 0.77 & 0.76 & 0.77 & 0 & 0 &  \\
J162437.86+345723.8 & 10.764 & 1.19 & 0.200 & 0.072 & 1.28 & 1.71 & 0.75 & 0 & 1 & S \\
J165424.59+241318.7 & 12.774 & 1.19 & 0.351 & 0.099 & 0.96 & 2.31 & 0.63 & 1 & 3 & B \\
    \hline
  \end{tabular}
\end{table*}

Shown in Figure 2 is a plot of the depth produced by orbiting extra-solar
planets of radii 0.5, 1.0, and 1.5 $R_J$ as a function of stellar radius.
The transit candidates from Table 3 are shown on the plot; the surviving
candidates as solid 5-pointed stars and the rejected candidates as open
circles. The candidate host stars are predominantly F--G--K stars as expected
of the spectral distribution amongst field stars, hence there does not appear
to be a significant bias towards early or late-type stars. Therefore, although
bloated gas-giant planets transiting late-type stars can result in transit
depths of $\sim 25$\%, these kinds of detections from wide-field surveys such
as SuperWASP will be quite rare.

\begin{figure}
  \includegraphics[angle=270,width=8.2cm]{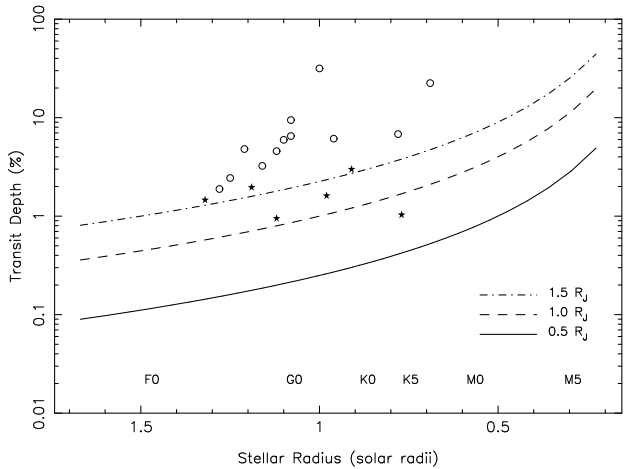}
  \caption{The transit depth for planets of radii 0.5, 1.0, and 1.5 $R_J$
    as a function of stellar radius. The rejected transit candidates from
    Table 3 are shown on the diagram as open circles and the final accepted
    candidates are shown as solid 5-pointed stars.}
\end{figure}

\subsection{Transit Candidates}

Presented here is a brief discussion for each of the six surviving candidates
from the third stage of the transit sifting process. The de-trended and
phase-folded lightcurves of these candidates along with their respective BLS
periodograms are shown in Figures 3 and 4. These figures also include
phase-binned average lightcurves weighted by $1/\sigma_i^2$ where $\sigma_i^2$
is the total estimated variance on each datapoint.

\begin{table*}
  \caption{The final list of candidates that passed all of the selection
    criteria, including the errors on the major fit parameters.}
  \begin{tabular}{@{}ccccccccccc}
    \hline
    SWASP ID    & Period & Depth & Duration & Epoch        \\
    (1SWASP$+$) & (days) & (mag) & (hours)  & (2450000$+$) \\
    \hline
J115718.66+261906.1 & $1.226804 \pm 0.000128$ & $0.0170 \pm 0.0014$ & $2.784 \pm 0.006$ & $3128.1643 \pm 0.0031$\\
J130322.00+350525.4 & $2.674207 \pm 0.000203$ & $0.0179 \pm 0.0010$ & $3.264 \pm 0.005$ & $3127.3608 \pm 0.0025$\\
J152131.01+213521.3 & $1.338018 \pm 0.000041$ & $0.0259 \pm 0.0010$ & $2.952 \pm 0.003$ & $3127.7908 \pm 0.0017$\\
J160211.83+281010.4 & $3.941554 \pm 0.000079$ & $0.0238 \pm 0.0007$ & $2.352 \pm 0.003$ & $3127.0391 \pm 0.0012$\\
J160242.43+290850.1 & $1.304693 \pm 0.000014$ & $0.0454 \pm 0.0010$ & $1.872 \pm 0.001$ & $3127.2471 \pm 0.0007$\\
J161732.90+242119.0 & $1.453738 \pm 0.000030$ & $0.0157 \pm 0.0008$ & $1.440 \pm 0.002$ & $3127.6799 \pm 0.0012$\\
    \hline
  \end{tabular}
\end{table*}

\begin{figure*}
  \begin{center}
    \begin{tabular}{cc}
      \includegraphics[width=8.2cm]{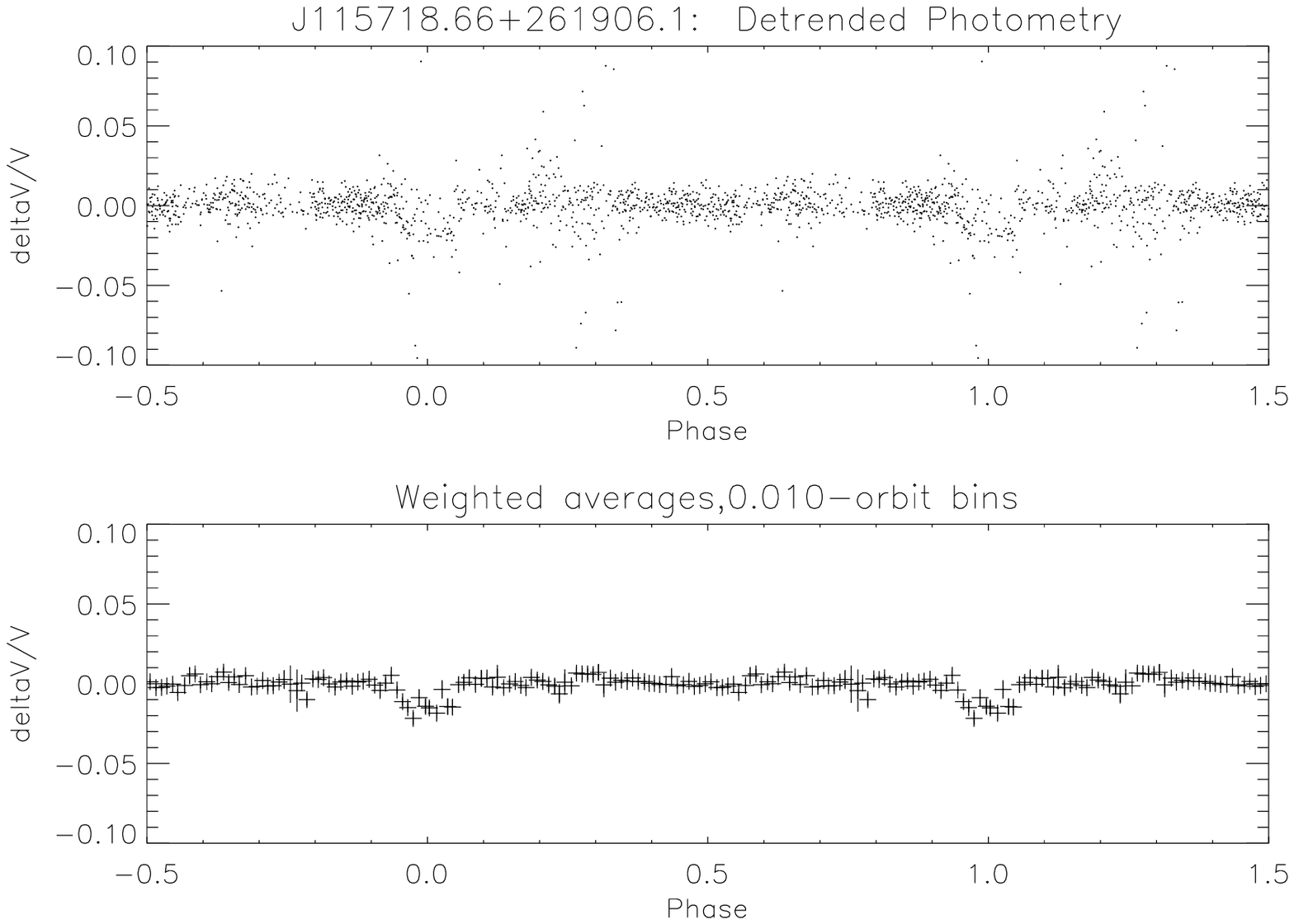} &
      \includegraphics[width=8.2cm]{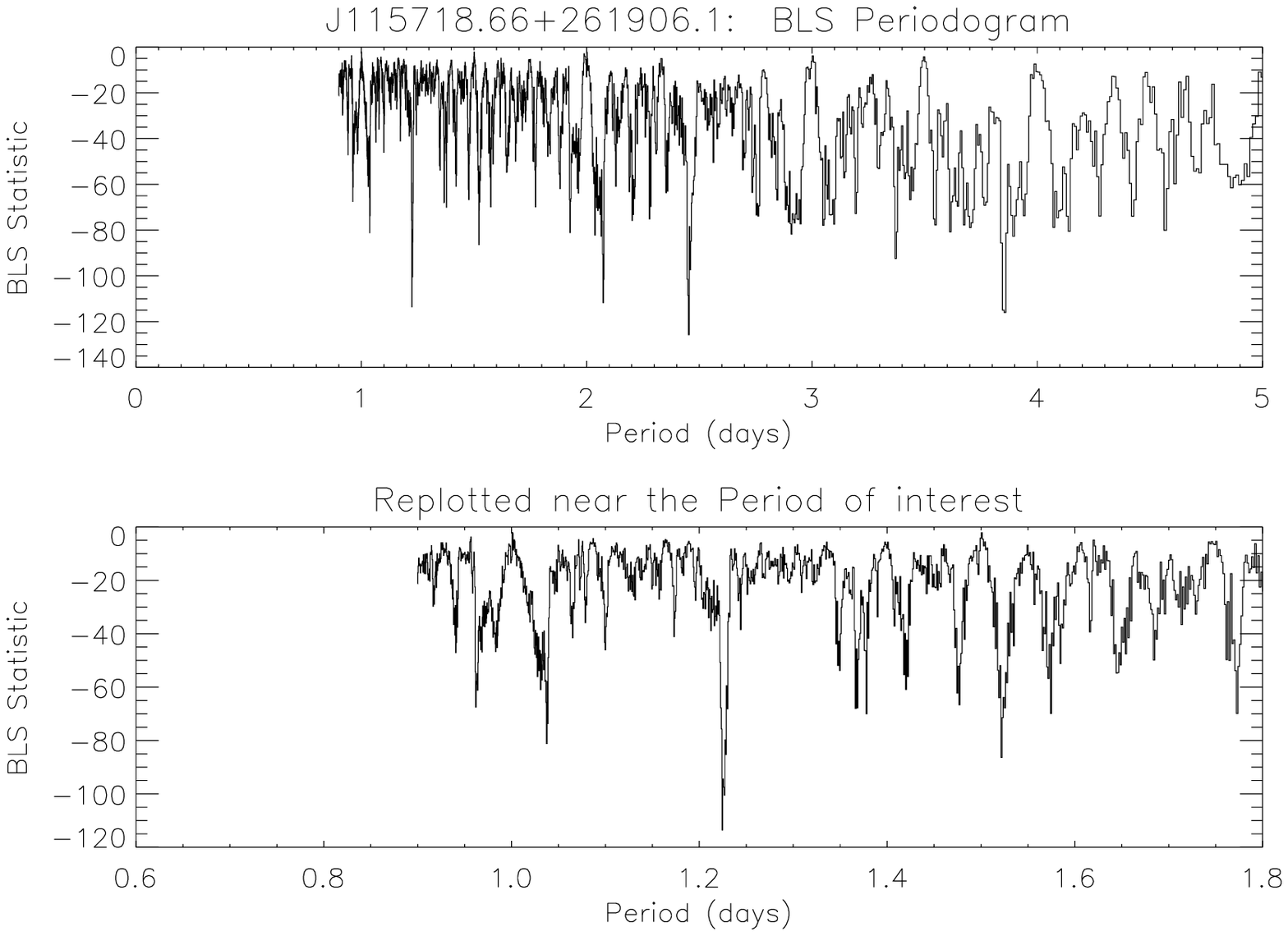} \\
      \\
      \includegraphics[width=8.2cm]{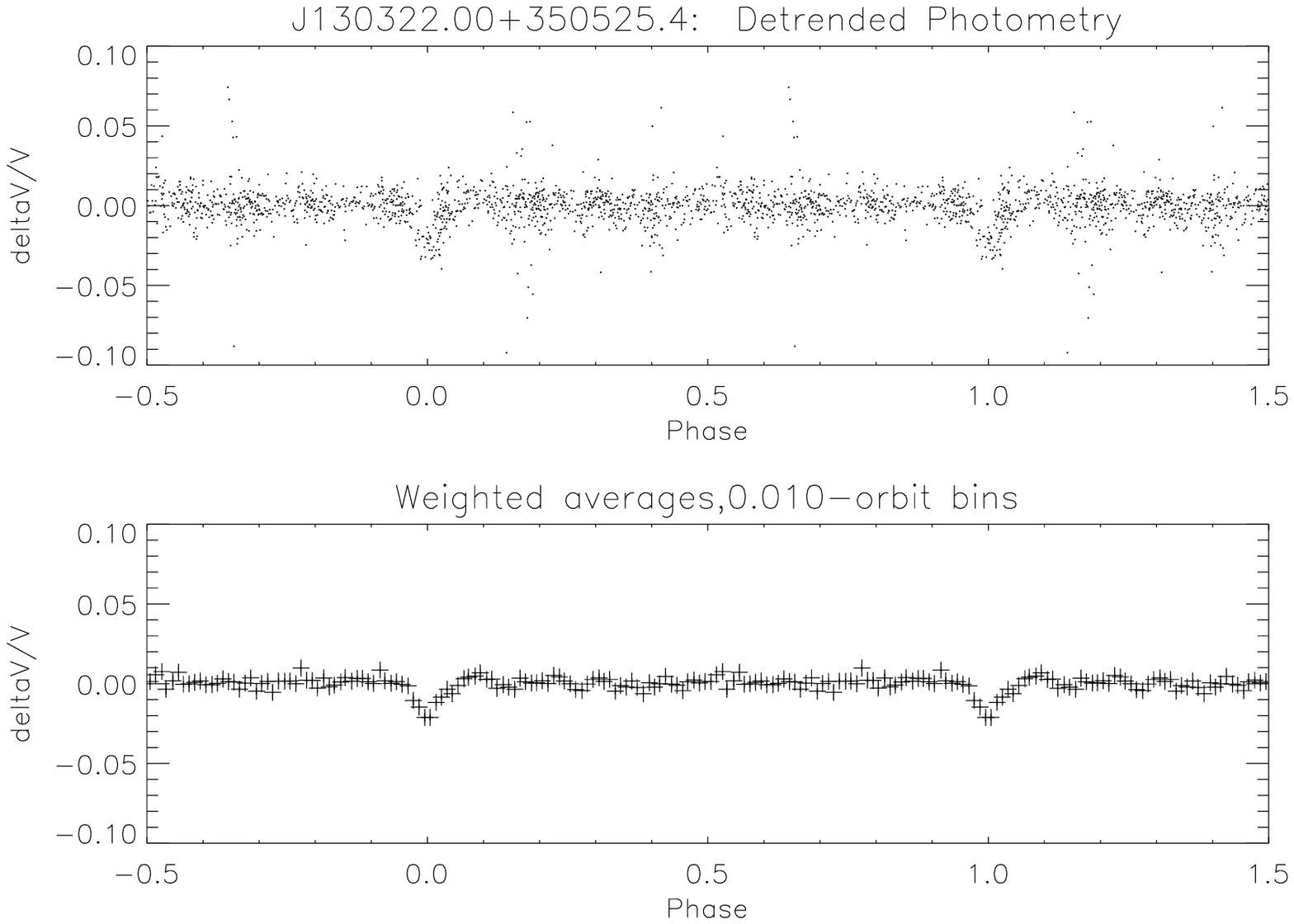} &
      \includegraphics[width=8.2cm]{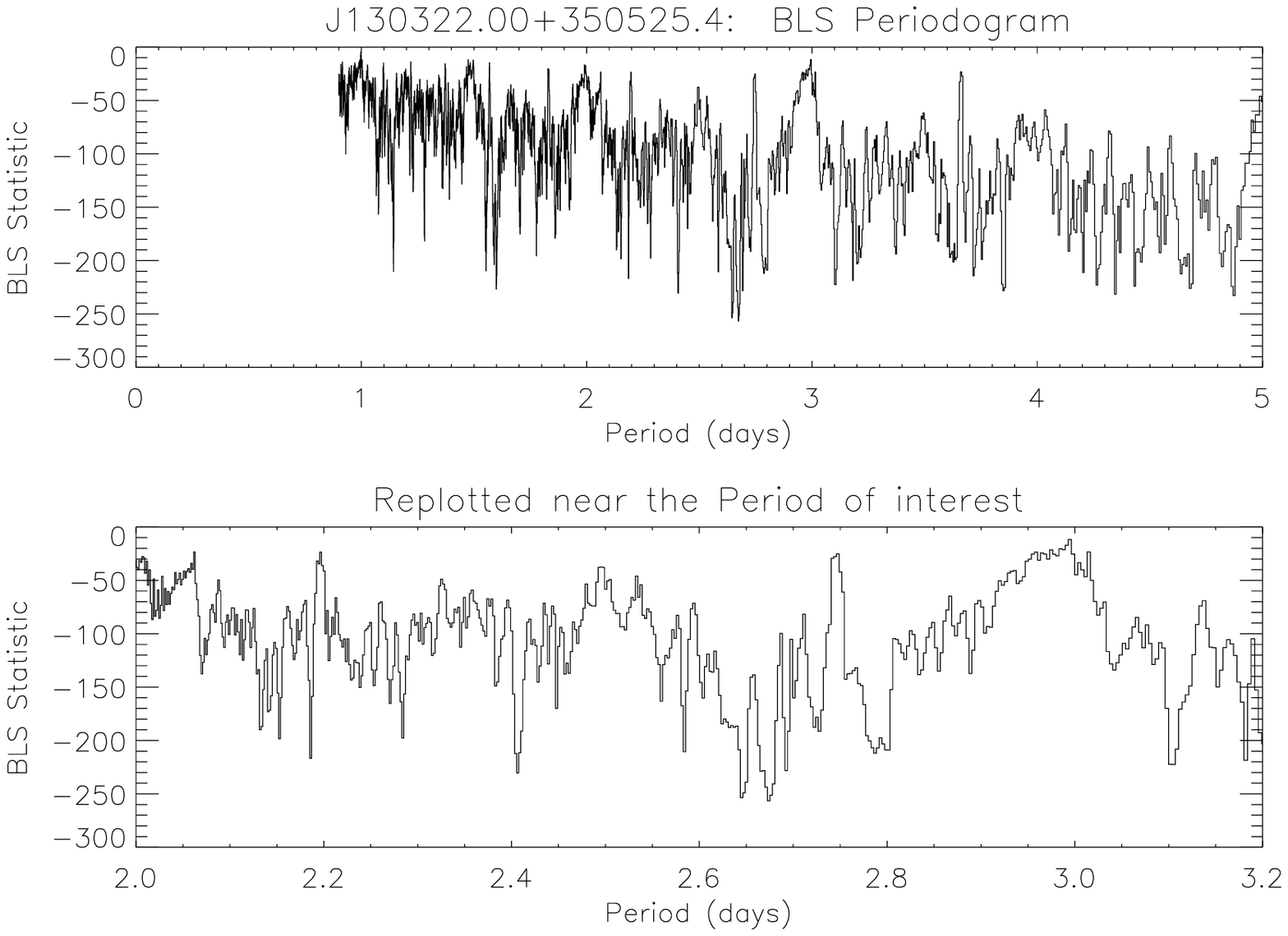} \\
      \\
      \includegraphics[width=8.2cm]{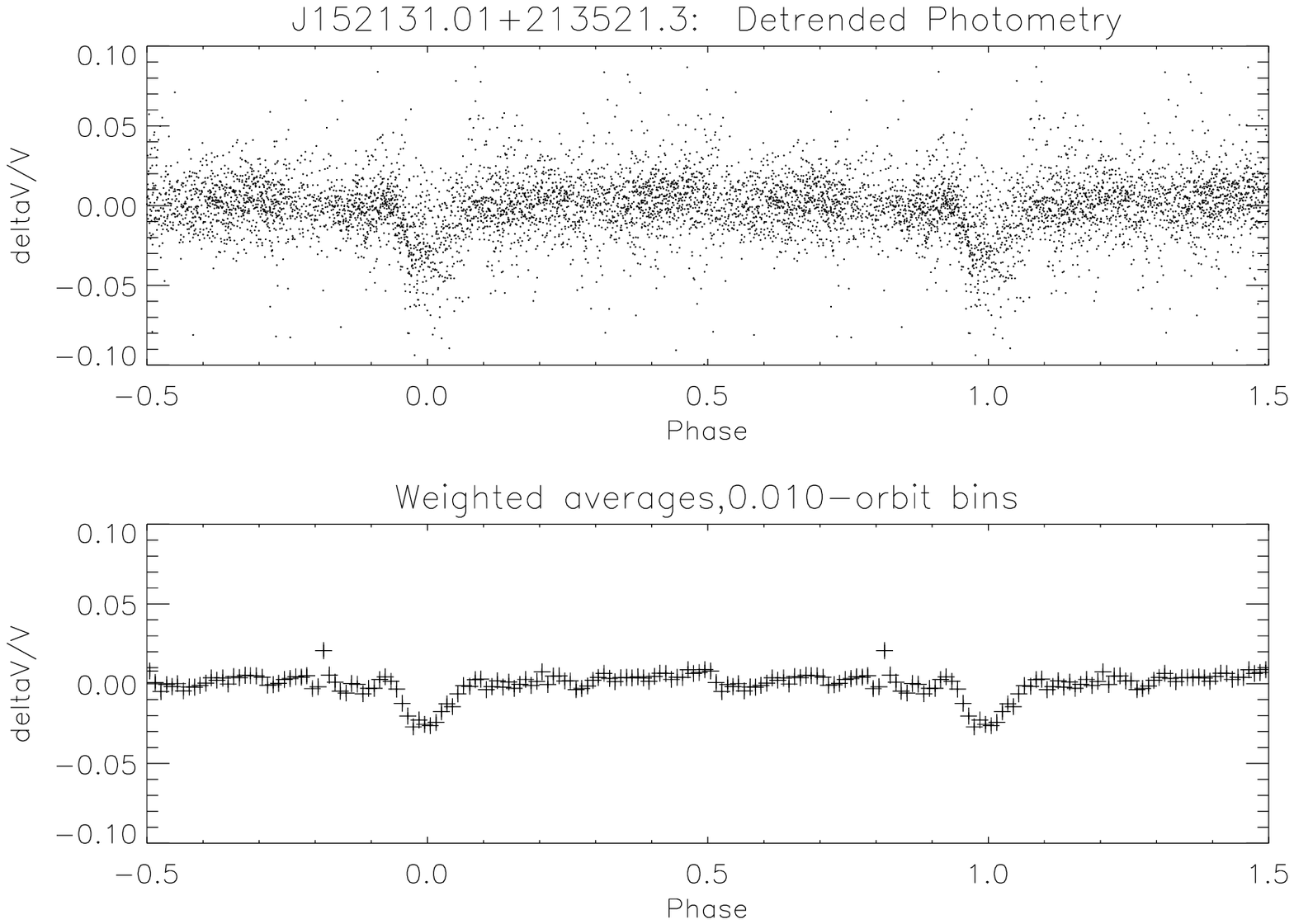} &
      \includegraphics[width=8.2cm]{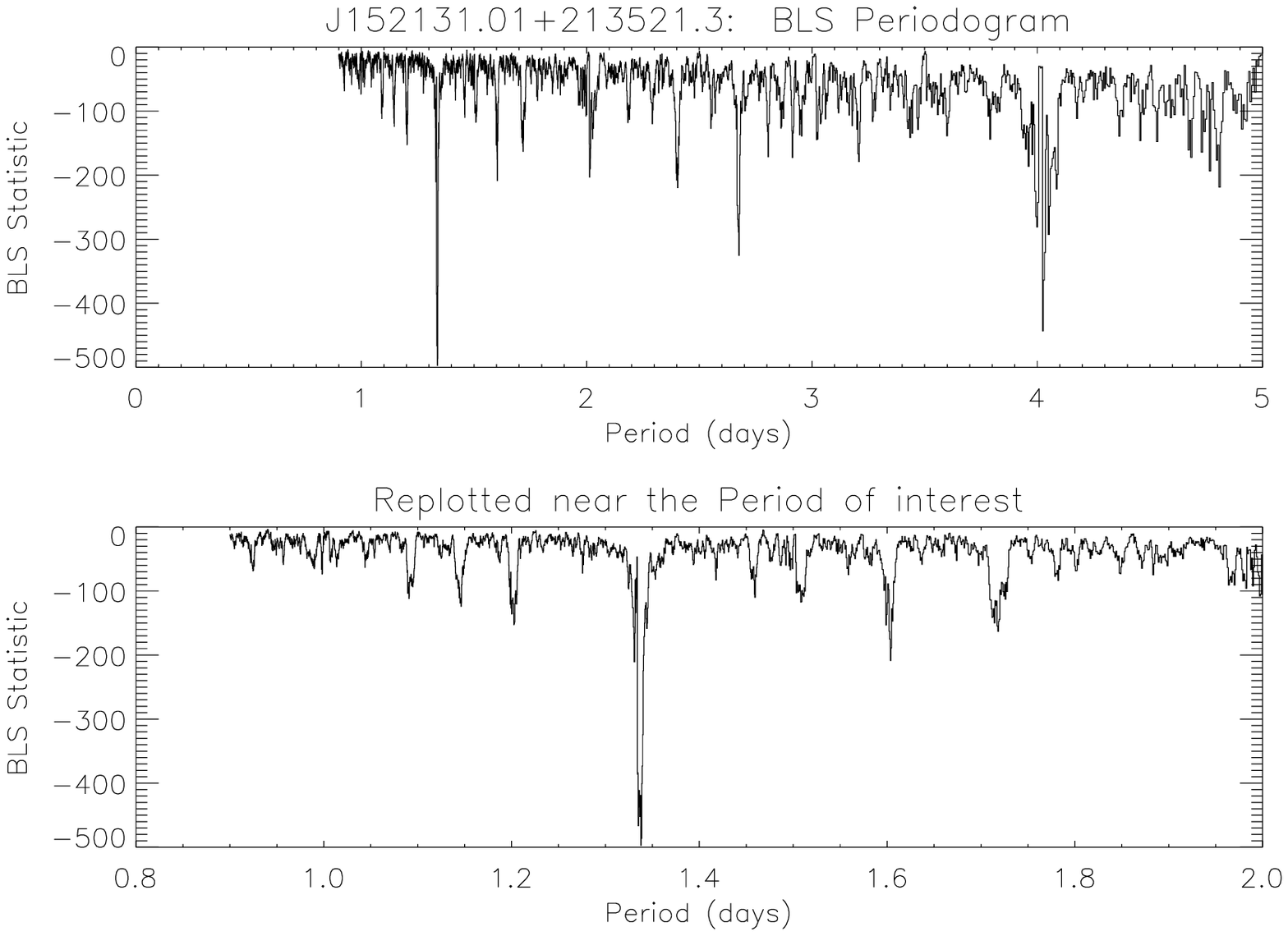} \\
    \end{tabular}
  \end{center}
  \caption{Final transit candidates 1--3, showing the unbinned and binned
    lightcurves (left) and the BLS periodograms (right).}
\end{figure*}

\begin{figure*}
  \begin{center}
    \begin{tabular}{cc}
      \includegraphics[width=8.2cm]{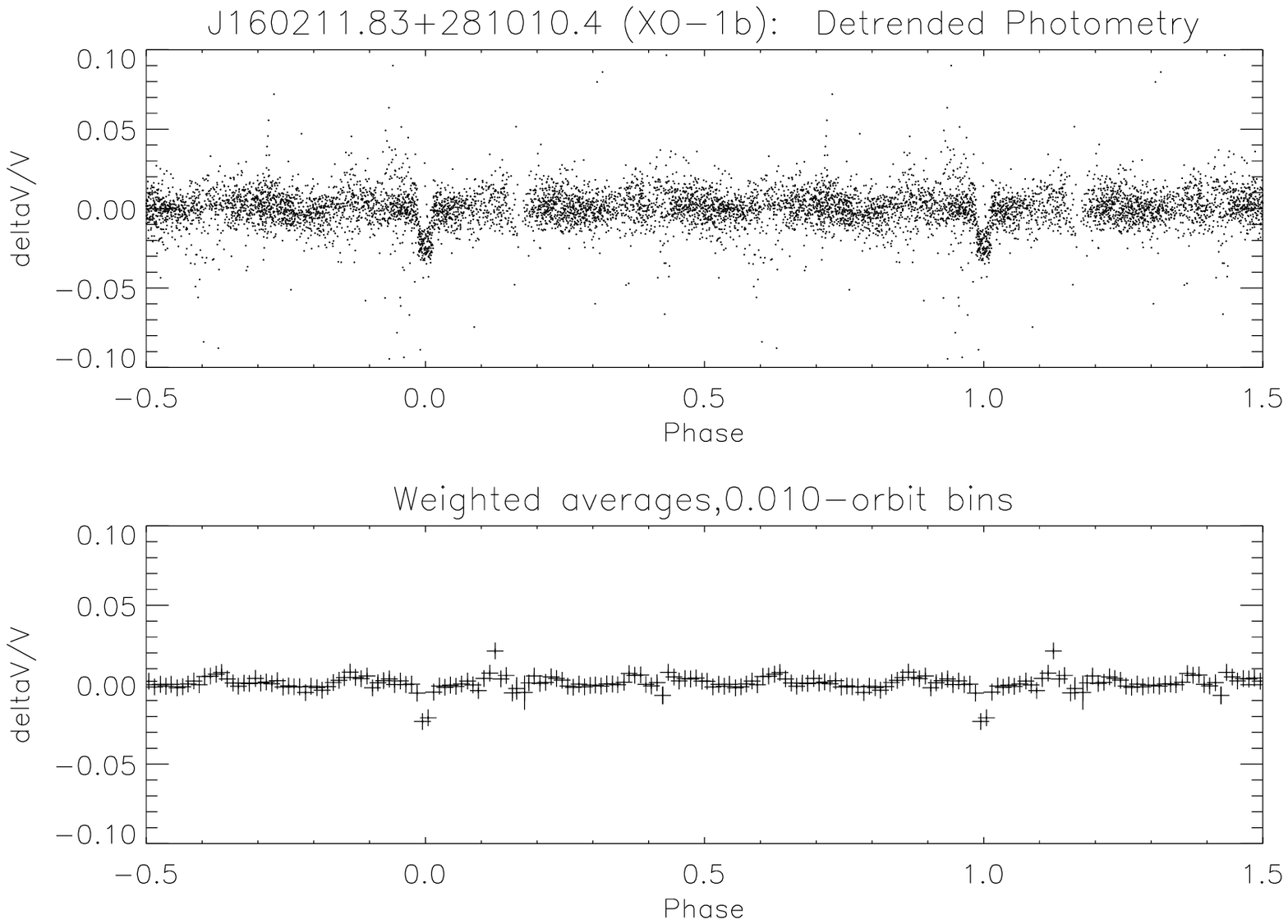} &
      \includegraphics[width=8.2cm]{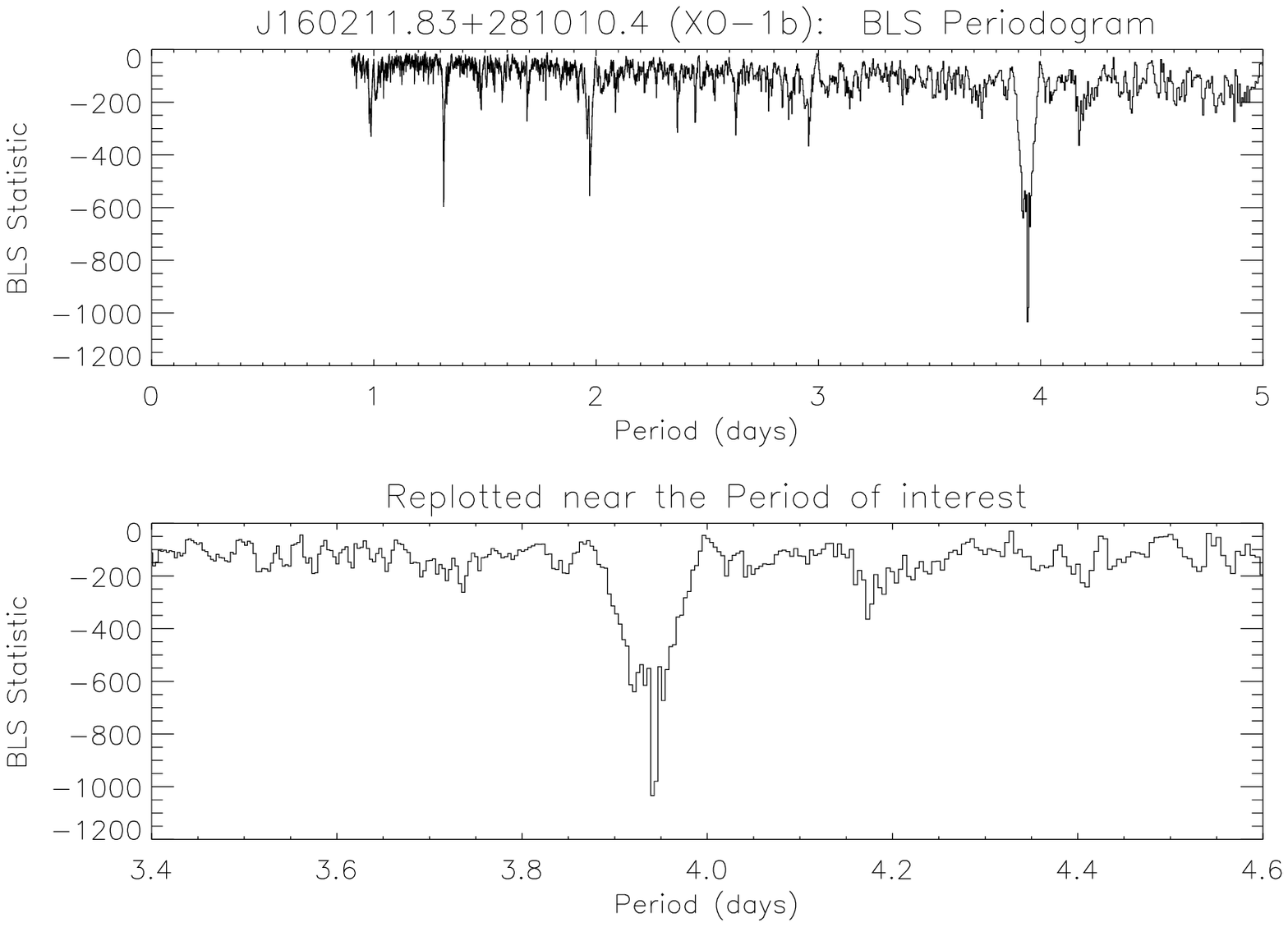} \\
      \\
      \includegraphics[width=8.2cm]{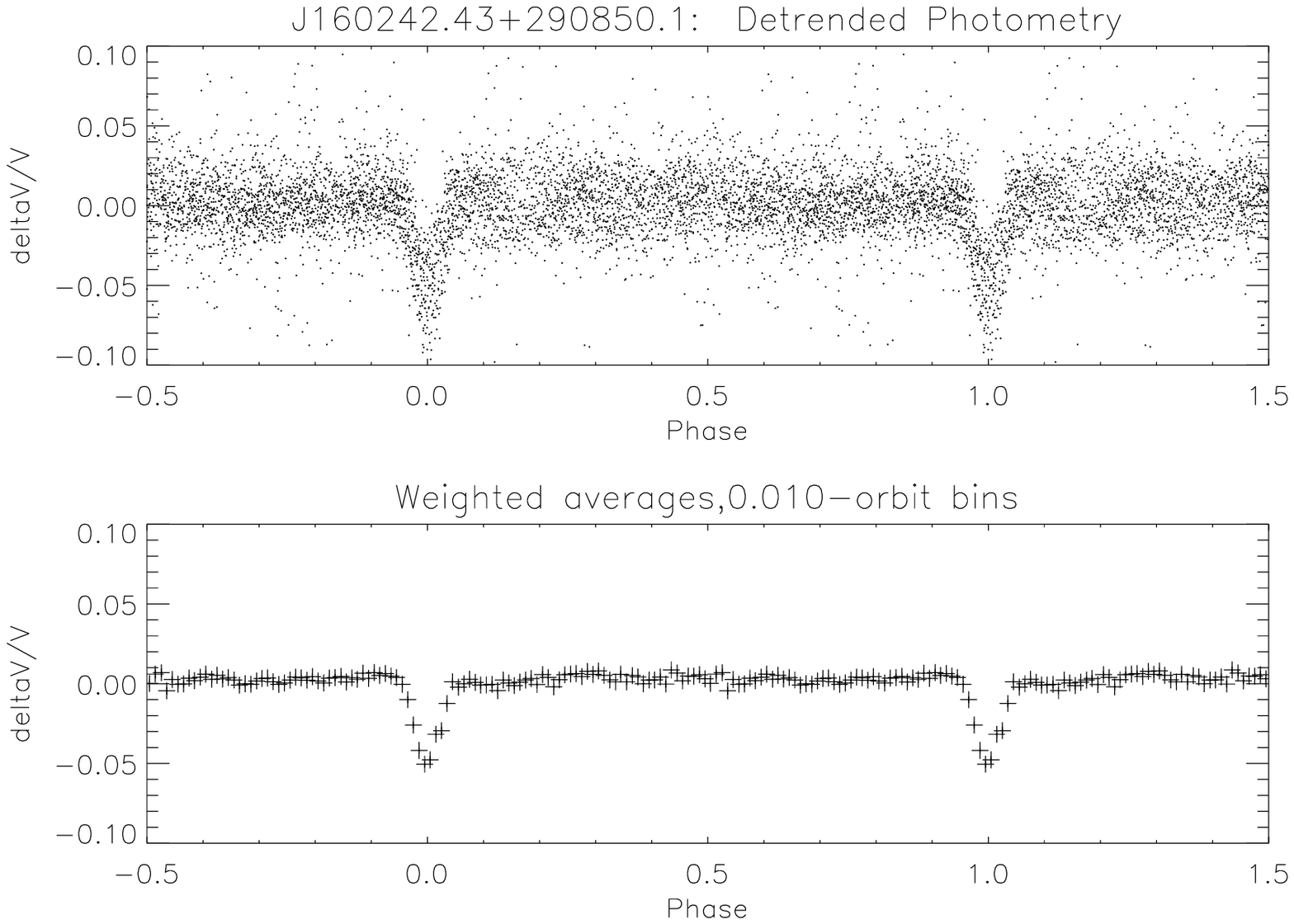} &
      \includegraphics[width=8.2cm]{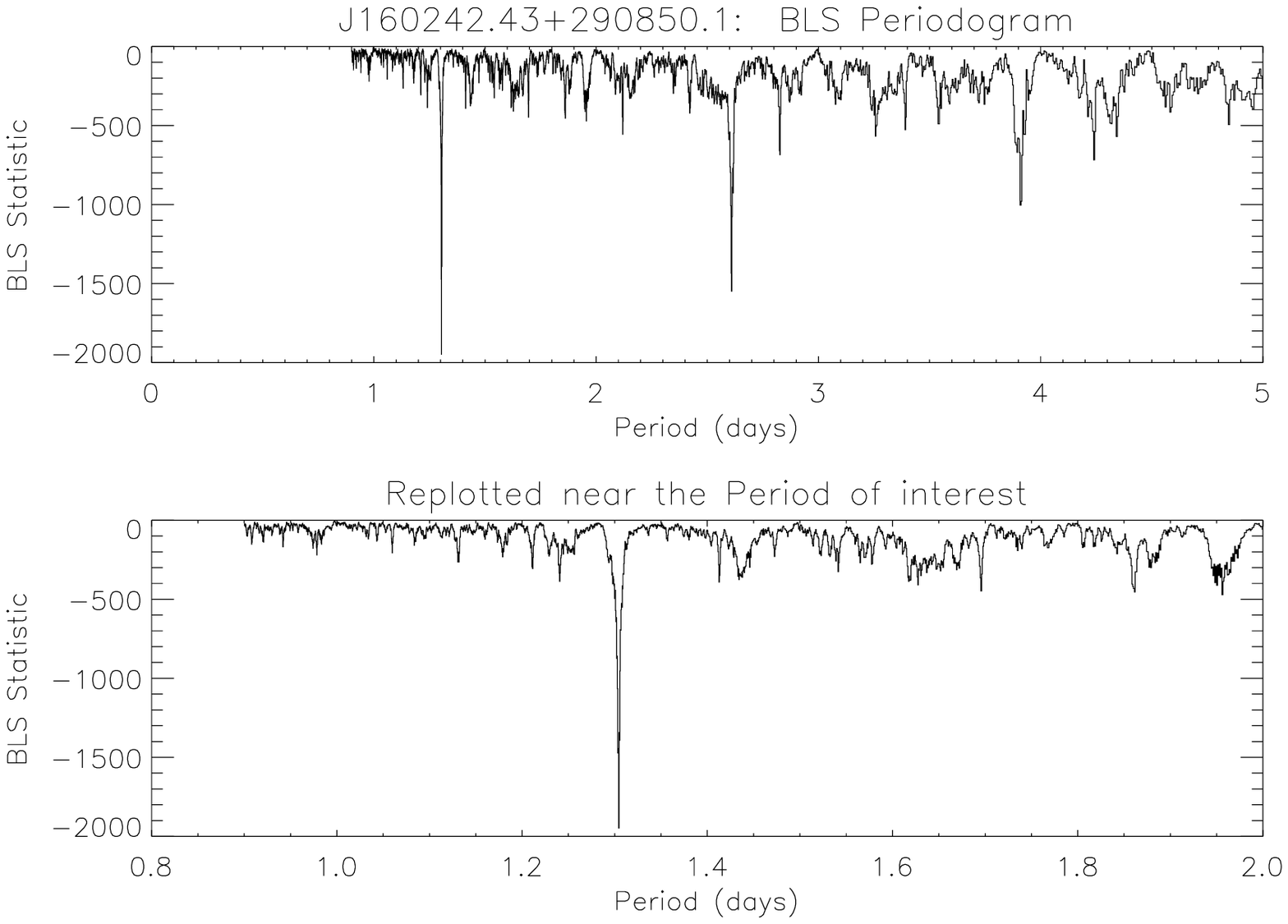} \\
      \\
      \includegraphics[width=8.2cm]{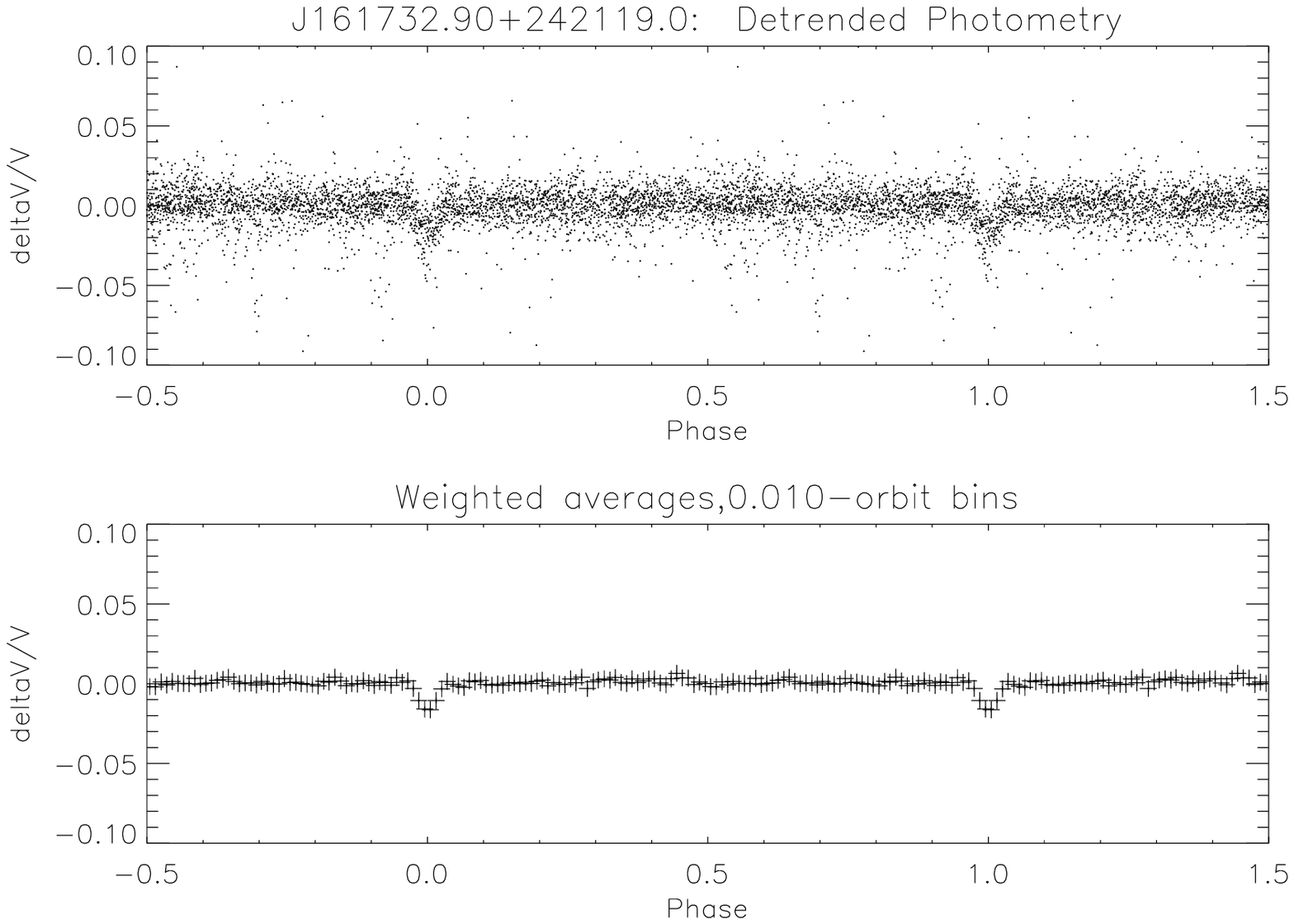} &
      \includegraphics[width=8.2cm]{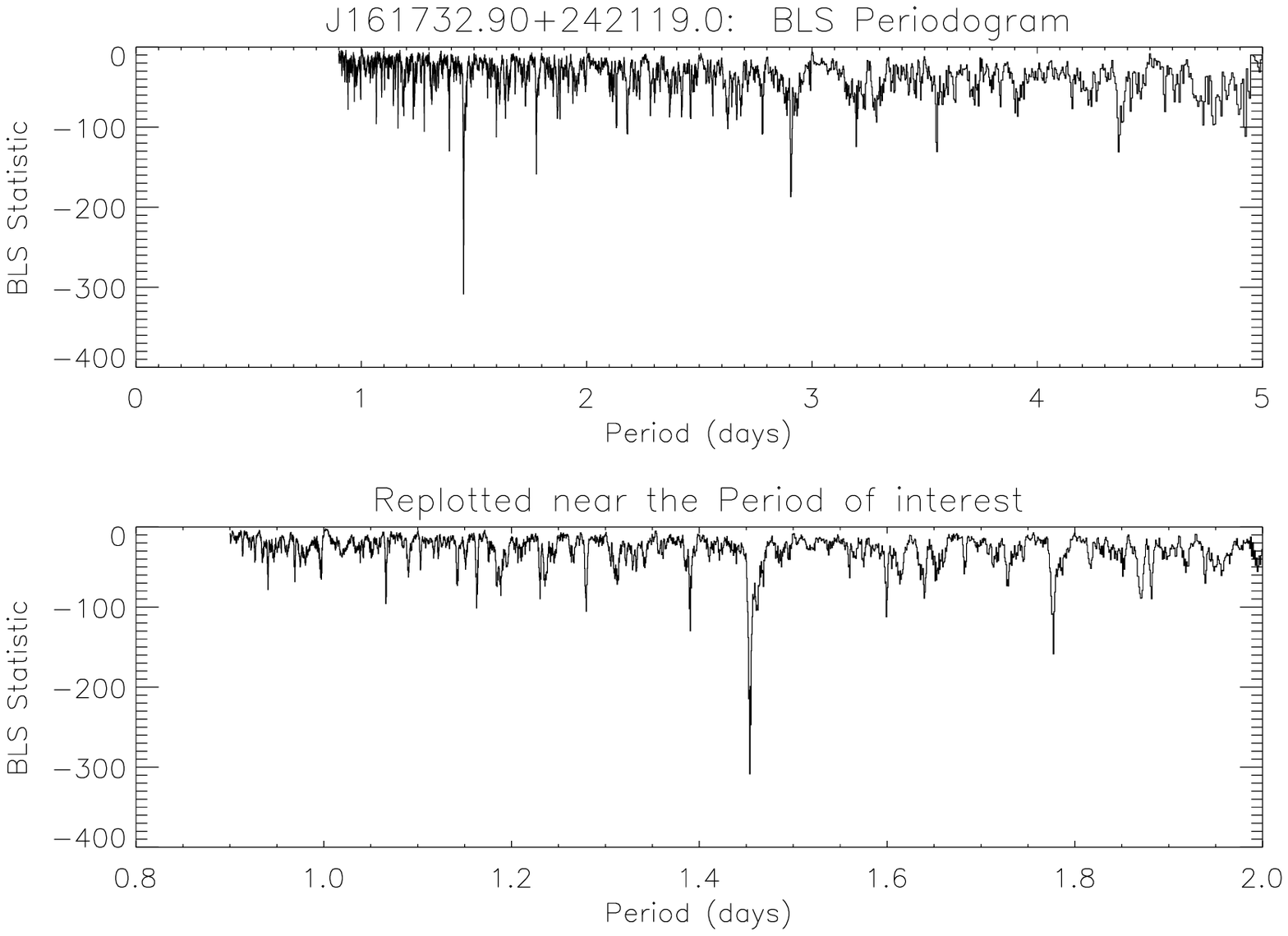} \\
    \end{tabular}
  \end{center}
  \caption{Final transit candidates 4--6, showing the unbinned and binned
    lightcurves (left) and the BLS periodograms (right).}
\end{figure*}

{\bf 1SWASP J115718.66+261906.1:} This candidate has 7 transits observed
with a period of $\sim 1.22$ days (the shortest period of the final
candidates) and a 2.78 hour duration. However, the periodogram reveals that
there are significant peaks at longer periods, especially at 2.45 days.
There is a relatively large amount of scatter in the lightcurve and the
signal-to-noise of the ellipsoidal variation $\mathrm{S/N}_{ellip}$ is
barely inside the rejection cutoff. It should also be noted that there are
two fainter objects within the 48\arcsec aperture. Even so, the lightcurve
exhibits a flat-bottomed transit and flat out-of transit behaviour. The
estimates of host spectral type and transit depth indicate a planet size of
1.55 $R_J$.

{\bf 1SWASP J130322.00+350525.4:} This is the brightest of the final
candidates although only 6 transits were observed during the 2004 campaign.
The period of $\sim 2.67$ days is well matched with the 3.26 hour duration
according to the value of $\eta_p$. The low value of $\mathrm{S/N}_{ellip}$
indicates that there is almost no ellipsoidal variation and is further
strengthened by the lack of any faint companions within the aperture. The
estimated size of the planet is 1.06 $R_J$. The transit appears to be
``V-shaped'' however so there is still a strong chance of this transit
candidate being a grazing eclipsing binary. Also, the lack of transits
cause the period determination to be relatively uncertain, as evidenced by
the lack of a strongly unique fourier power in the periodogram.

{\bf 1SWASP J152131.01+213521.3:} This candidate is the faintest to meet
all of the selection criteria. A total of 21 transits were observed however
resulting in a very strong peak in the periodogram at a period of $\sim
1.34$ days. The ellipsoidal variation is very low, the signal to red noise
ratio is very high, and the transit is flat-bottomed. The estimate of the
planetary radius in this case is 1.62 $R_J$, which is slightly above the
radius cutoff described in section 3.3 but the other positive attributes of
the candidate justified its inclusion in the final list.

{\bf 1SWASP J160211.83+281010.4:} This candidate is in fact the known
transiting planet XO-1b, which was discovered by \citet{mcc06}. It is
encouraging that this planet was detected with a high degree of confidence
(high $\Delta \chi^2$) by the detection algorithm, for which 11 transits
were observed. The VSI analysis detected no brighter or fainter companions
within the aperture. The orbital period of 3.94 days and estimated radius
of 1.21 $R_J$ are in good agreement with the values reported by
\citet{mcc06}. A more detailed analysis of the SuperWASP-N XO-1b data may
be found in \citet{wil06}.

{\bf 1SWASP J160242.43+290850.1:} This relatively faint candidate has 23
observed transits with a best-fit period of $\sim 1.30$ days and a duration
of 1.87 hours. Based upon the spectral type of the host star, the estimated
radius of the planet is 1.53 $R_J$. However, there is a fainter object
within the aperture and the transit is ``V-shaped'' which means that this
is possibly due to a grazing eclipsing binary rather than a true planetary
transit. Additionally, there is a second dominant peak in the periodogram
at $\sim 2.61$ days. Folding the data on this longer period reveals two
eclipses of slightly different depth. This further evidence indicates that
this candidate is most likely not due to a planet.

{\bf 1SWASP J161732.90+242119.0:} There are 16 transits observed for this
candidate with a period of $\sim 1.45$ days and a duration of 1.44 hours.
The ellipsoidal variation is extremely low and there are no detected faint
companions within the aperture. The large colour index of this star
indicates a late-K spectral type which leads to a relatively small
estimate for the planet radius of 0.76 $R_J$. The promising nature of this
candidate led to spectroscopic follow-up observations, described in
section 4.3.

\subsection{Follow-up of Transit Candidates}

The major source of transit mimics amongst candidates are eclipsing
binaries. These can be either grazing eclipsers of $\sim 1\%$ depth or
blended eclipsers contributing $\sim 1\%$ of light \citep{bro03}.
Wide-field surveys such as SuperWASP and WASP0 generally suffer from
heavily undersampled stellar profiles due to the large pixel sizes. In
most cases, eclipsing binaries can be excluded through photometric
analysis, the catalogue queries provided by VSI, or straightforward
multi-colour observations using a higher angular resolution telescope to
resolve blended objects.

The techniques described in section 3 were used on all the SuperWASP-N
2004 fields to construct a list of high priority candidates. Further
study of these candidates required precision radial velocity measurements
to test the planet hypothesis of the observed transit events. A sample of
the high priority candidates were subsequently observed using the SOPHIE
cross-dispersed echelle spectrograph on the 1.93m telescope at the
Observatoire de Haute-Provence \citep{col07a}. SOPHIE achieved first light
on 31st July, 2006 and so the follow-up campaign undertaken by the
SuperWASP consortium was amongst the first science applications of the
instrument.

Amongst these candidates observed during these runs was the last of the
candidates shown in Table 4: 1SWASP J161732.90+242119.0. A single
narrow-lined cross-correlation function (CCF) was observed for each of
the spectra obtained. Evidence of pressure broadening was seen in the
Na\,{\sevensize I}\,$D$ and Mg\,{\sevensize I}\,$D$ lines with slight
asymmetry and red-shifted emission in the H$\alpha$ line. A radial
velocity variation of only a few m/s determined from three spectra were
used to conlude that there is no significant evidence of radial velocity
induced behaviour due to the presence of a planet. Hence, this candidate
has been ruled out as a transiting planet from the SOPHIE observations.
The remainder of the candidates have been assigned priorities and await
further observations.

\subsection{Rejected Candidates}

Presented here is a brief discussion for a subset of the rejected
candidates shown in Tables 2 and 3. The purpose of this discussion is to
highlight the quality of these candidates and hence the potential for
false-alarms being needlessly observed with spectroscopic follow-up. The
lightcurves and BLS periodigrams for these six examples are shown in
Figures 5 and 6. These exhibit strong transit-like signatures but the
information available for the majority of the rejected candidates from
stages two and three of the sifting process indicated that the size of
the secondary is too large to be a planetary companion.

\begin{figure*}
  \begin{center}
    \begin{tabular}{cc}
      \includegraphics[width=8.2cm]{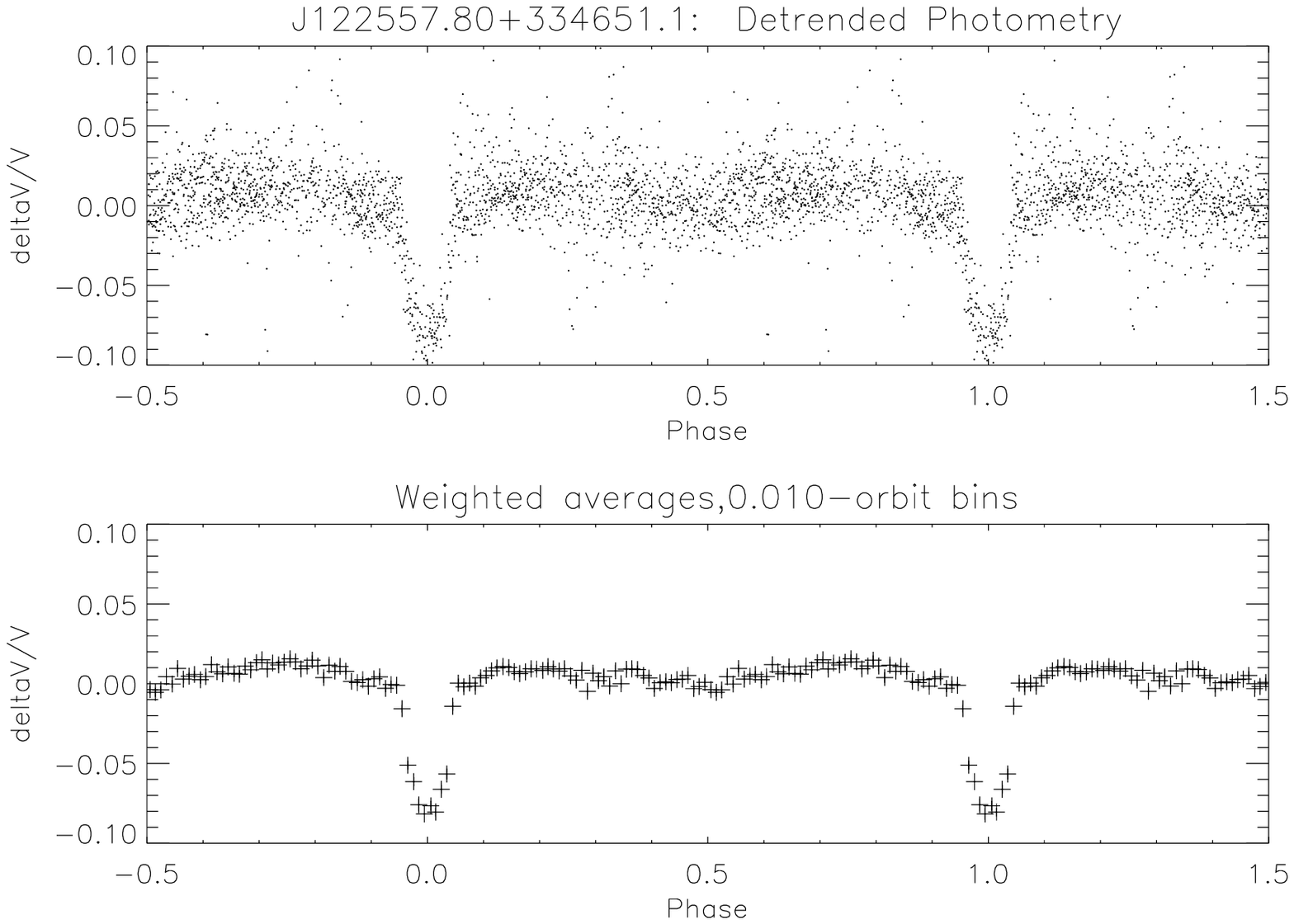} &
      \includegraphics[width=8.2cm]{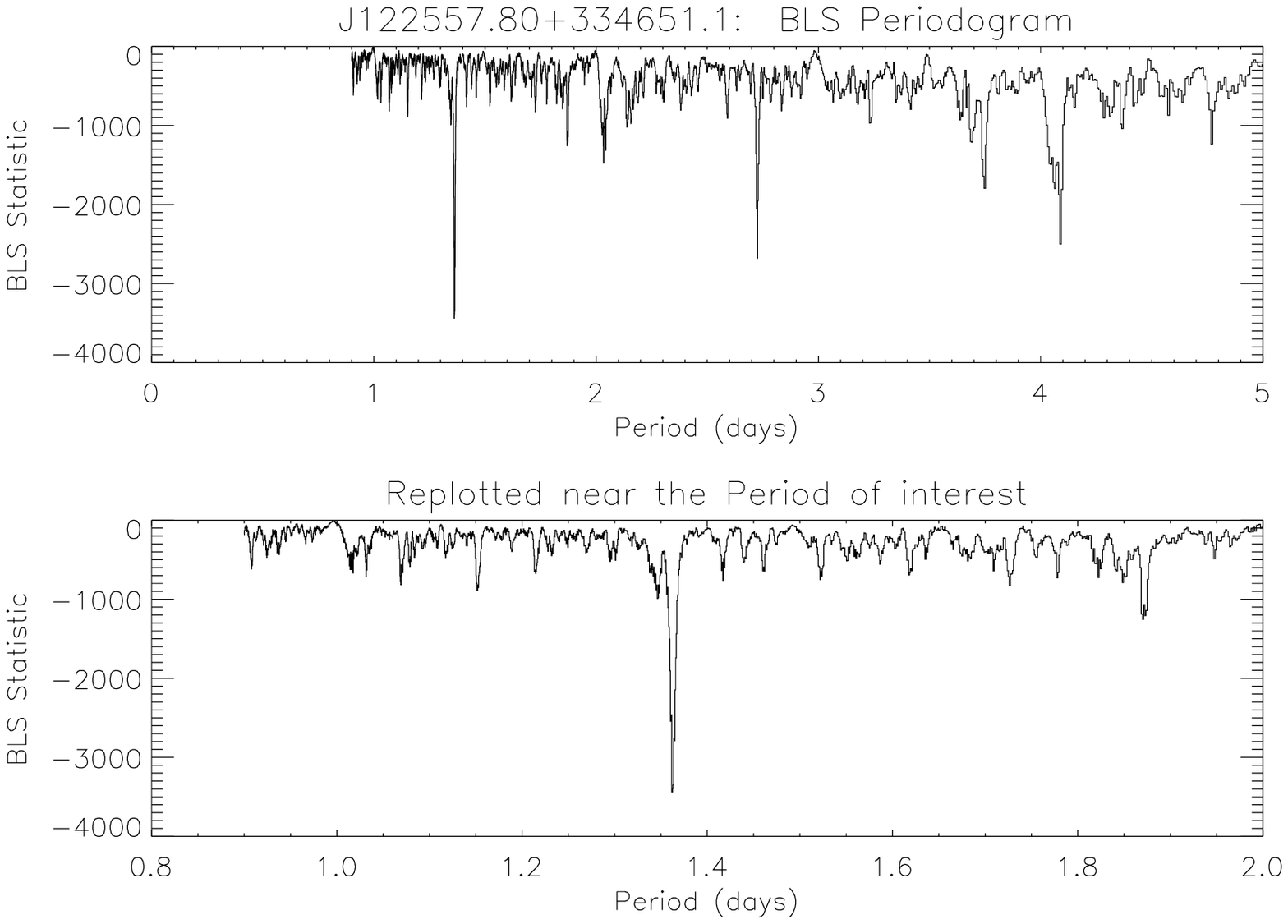} \\
      \\
      \includegraphics[width=8.2cm]{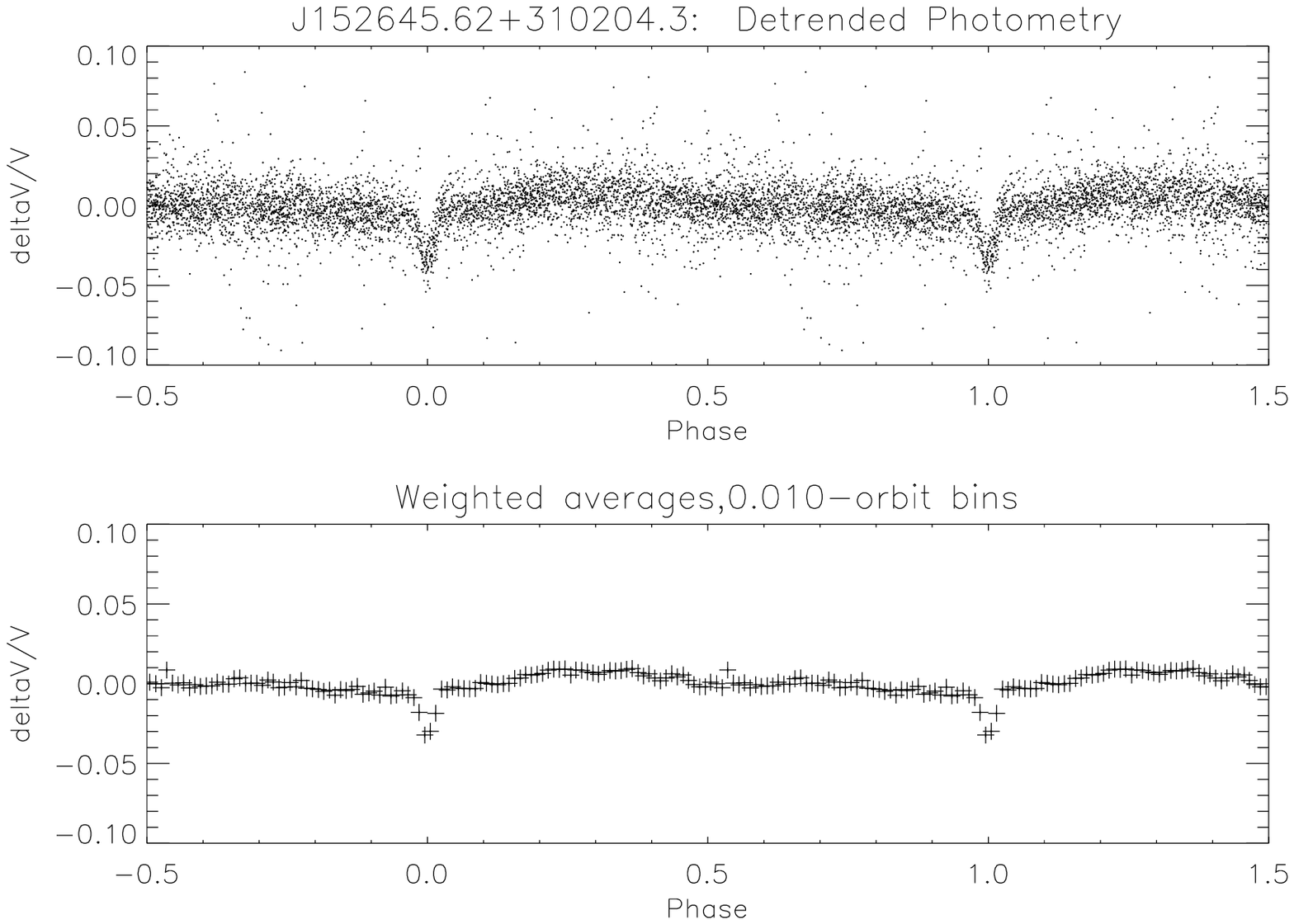} &
      \includegraphics[width=8.2cm]{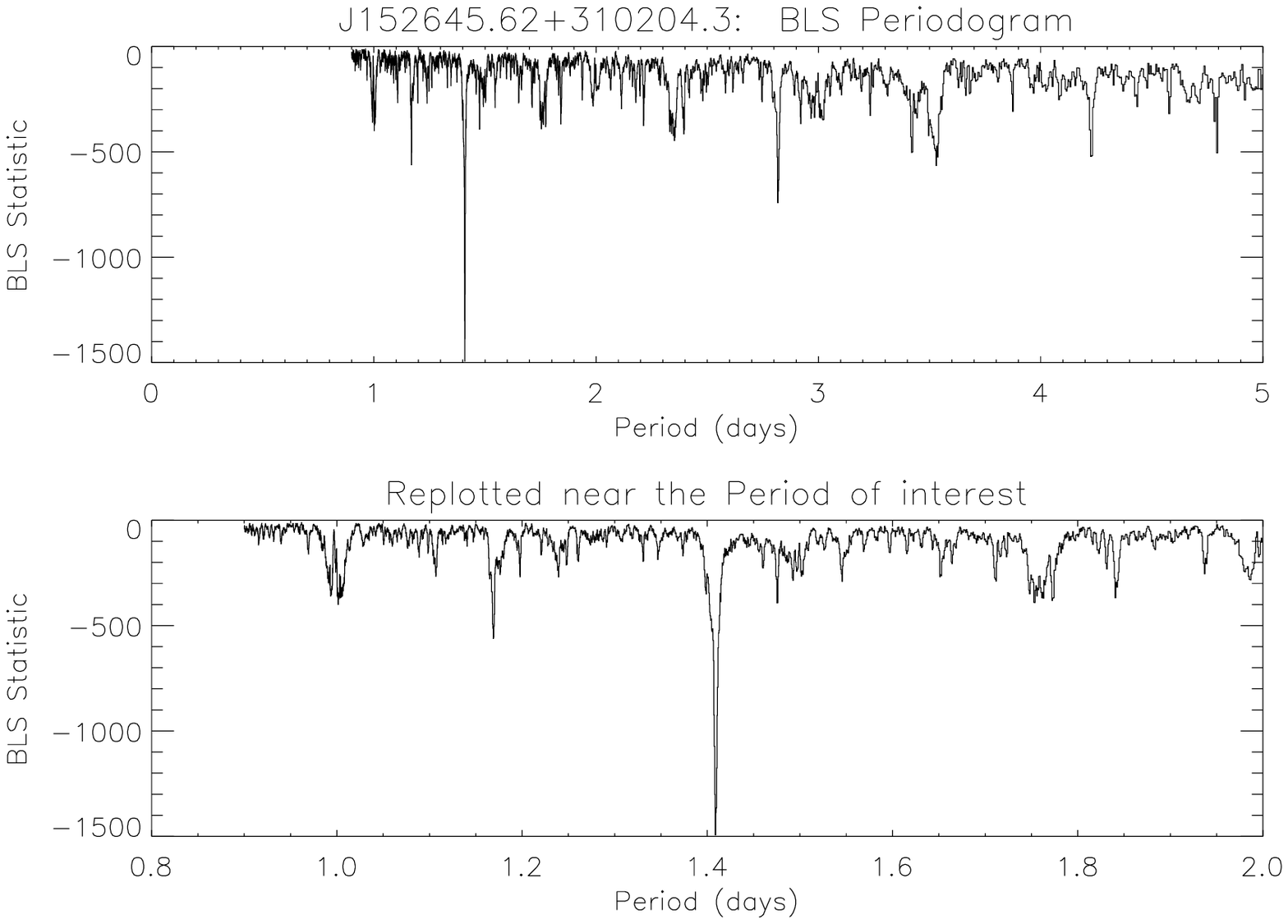} \\
      \\
      \includegraphics[width=8.2cm]{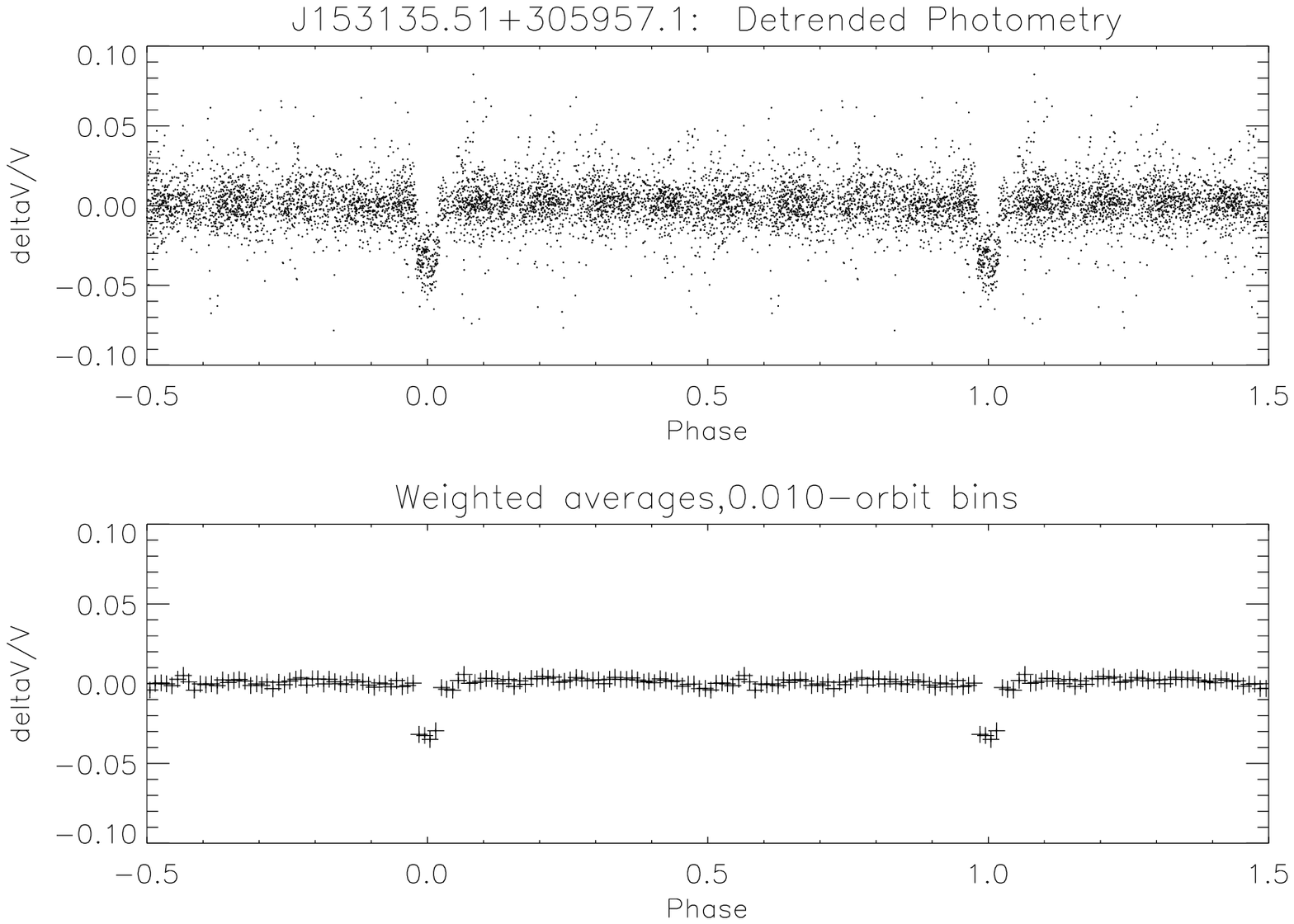} &
      \includegraphics[width=8.2cm]{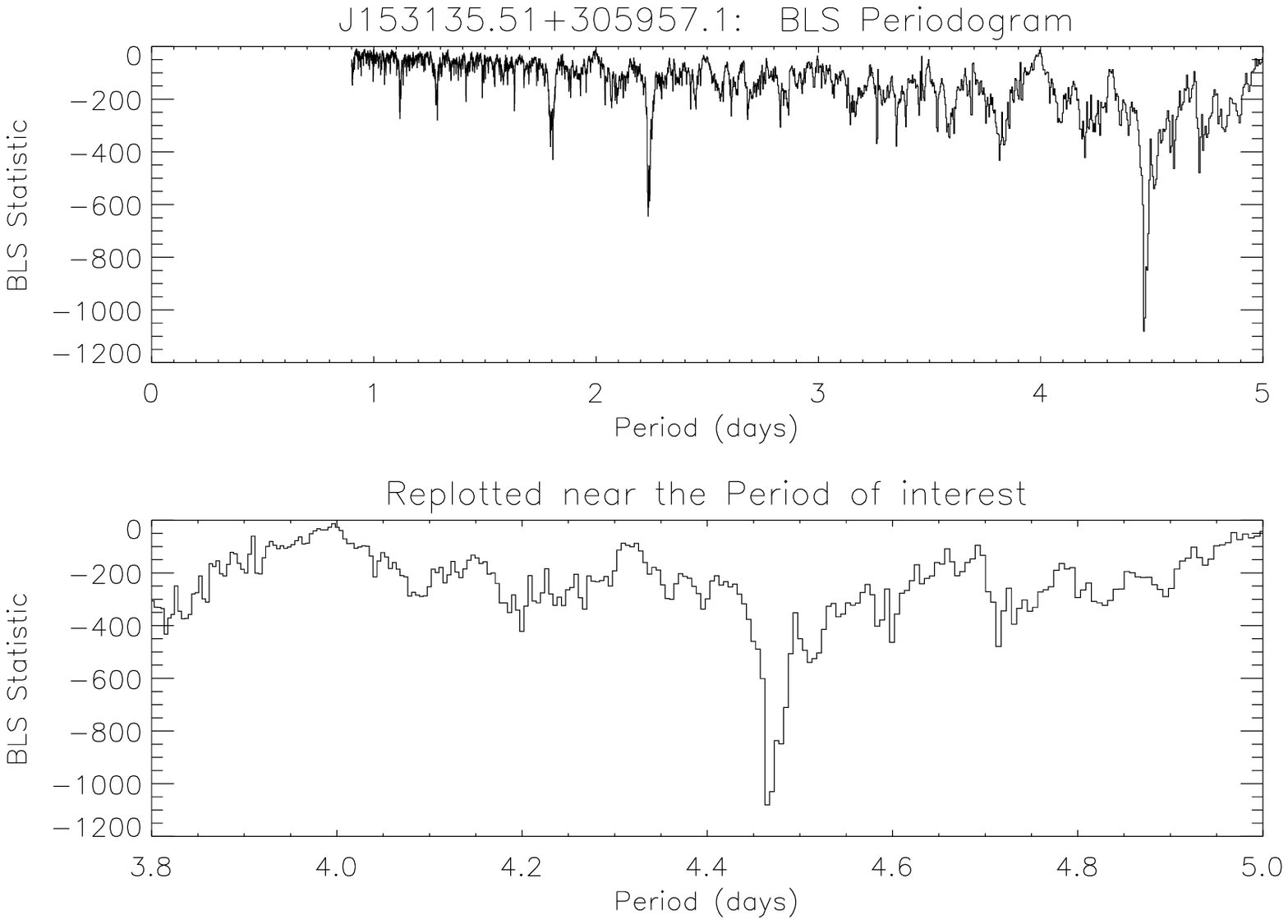} \\
    \end{tabular}
  \end{center}
  \caption{Example rejected candidates, showing the unbinned and binned
    lightcurves (left) and the BLS periodograms (right).}
\end{figure*}

\begin{figure*}
  \begin{center}
    \begin{tabular}{cc}
      \includegraphics[width=8.2cm]{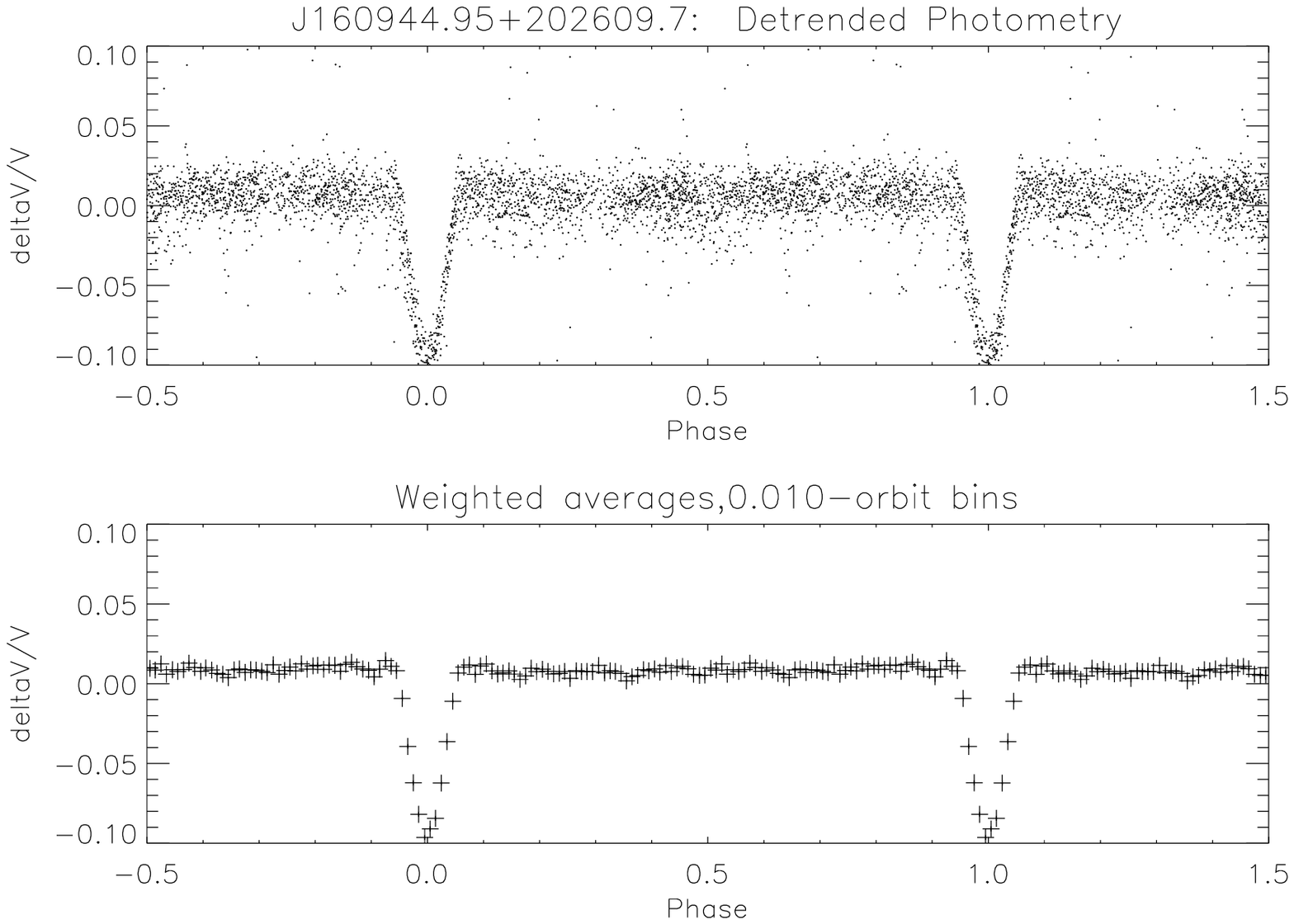} &
      \includegraphics[width=8.2cm]{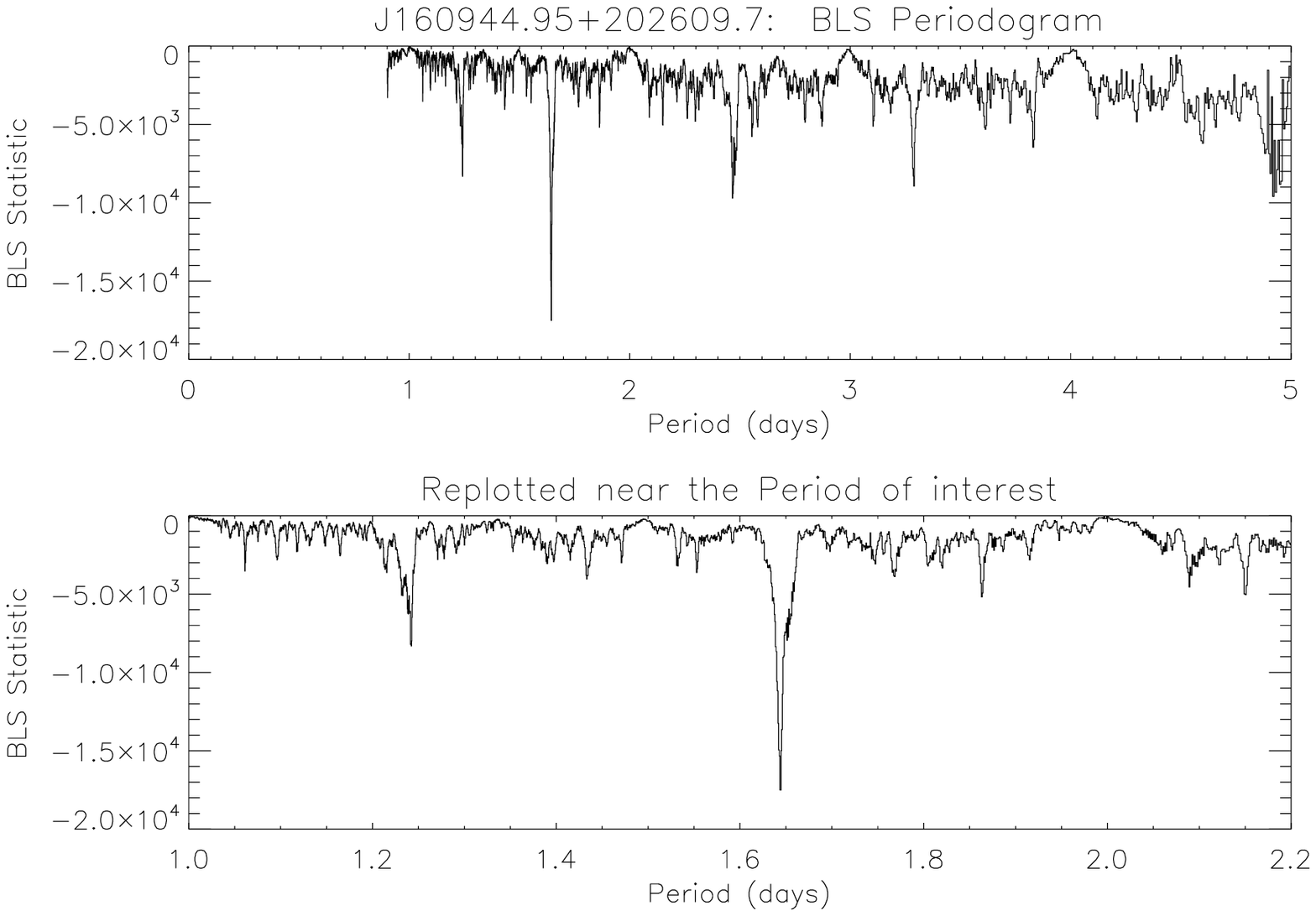} \\
      \\
      \includegraphics[width=8.2cm]{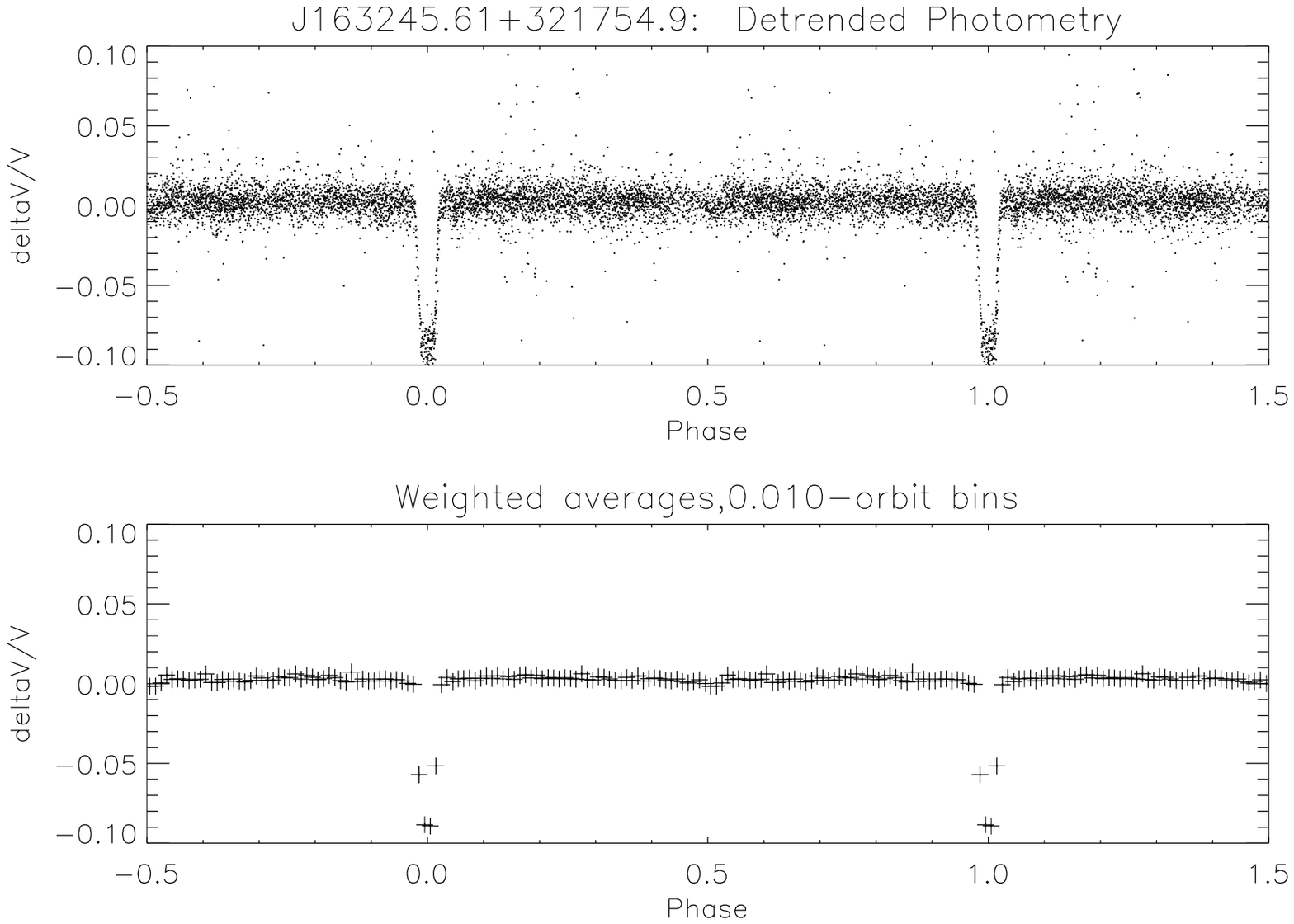} &
      \includegraphics[width=8.2cm]{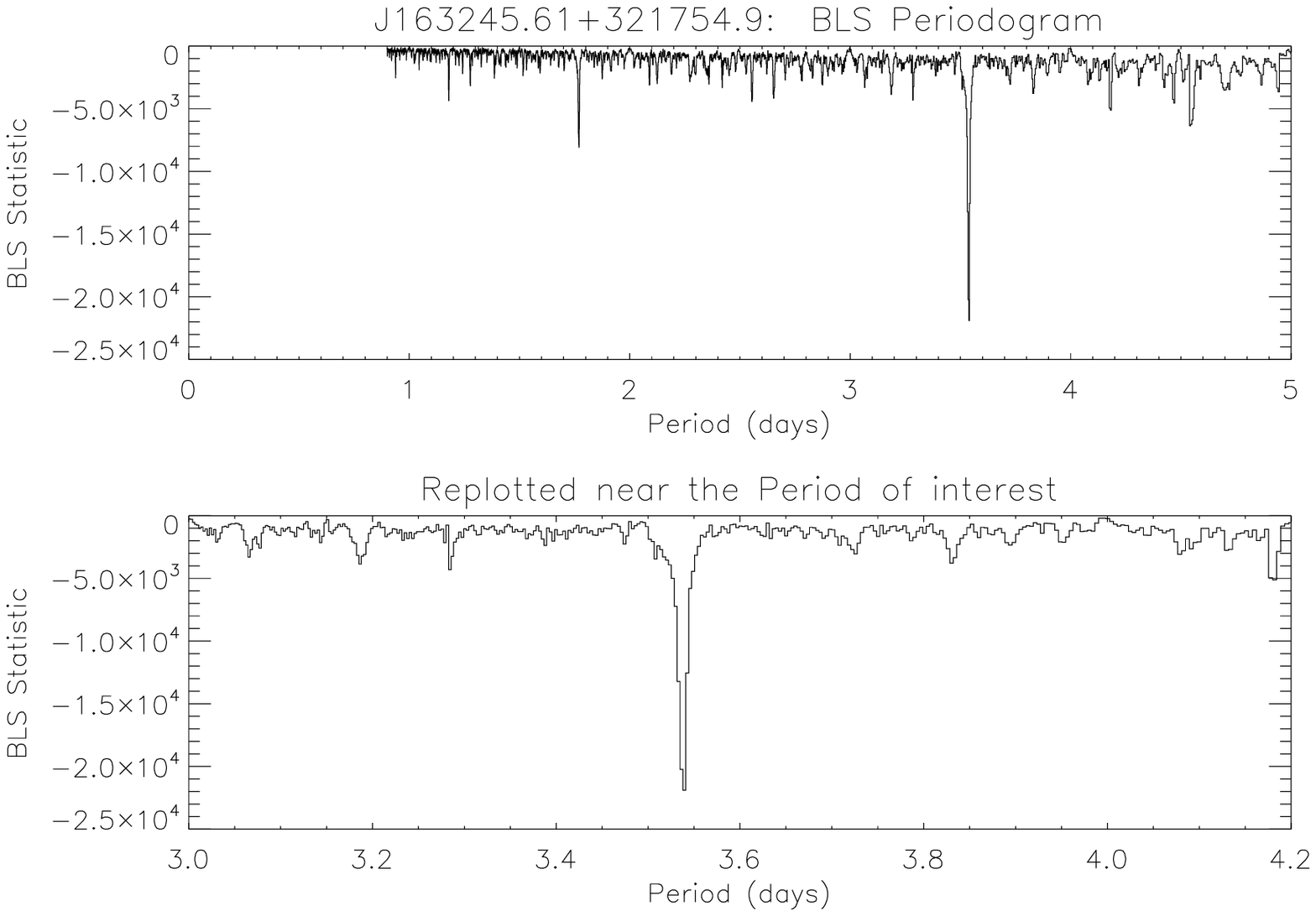} \\
      \\
      \includegraphics[width=8.2cm]{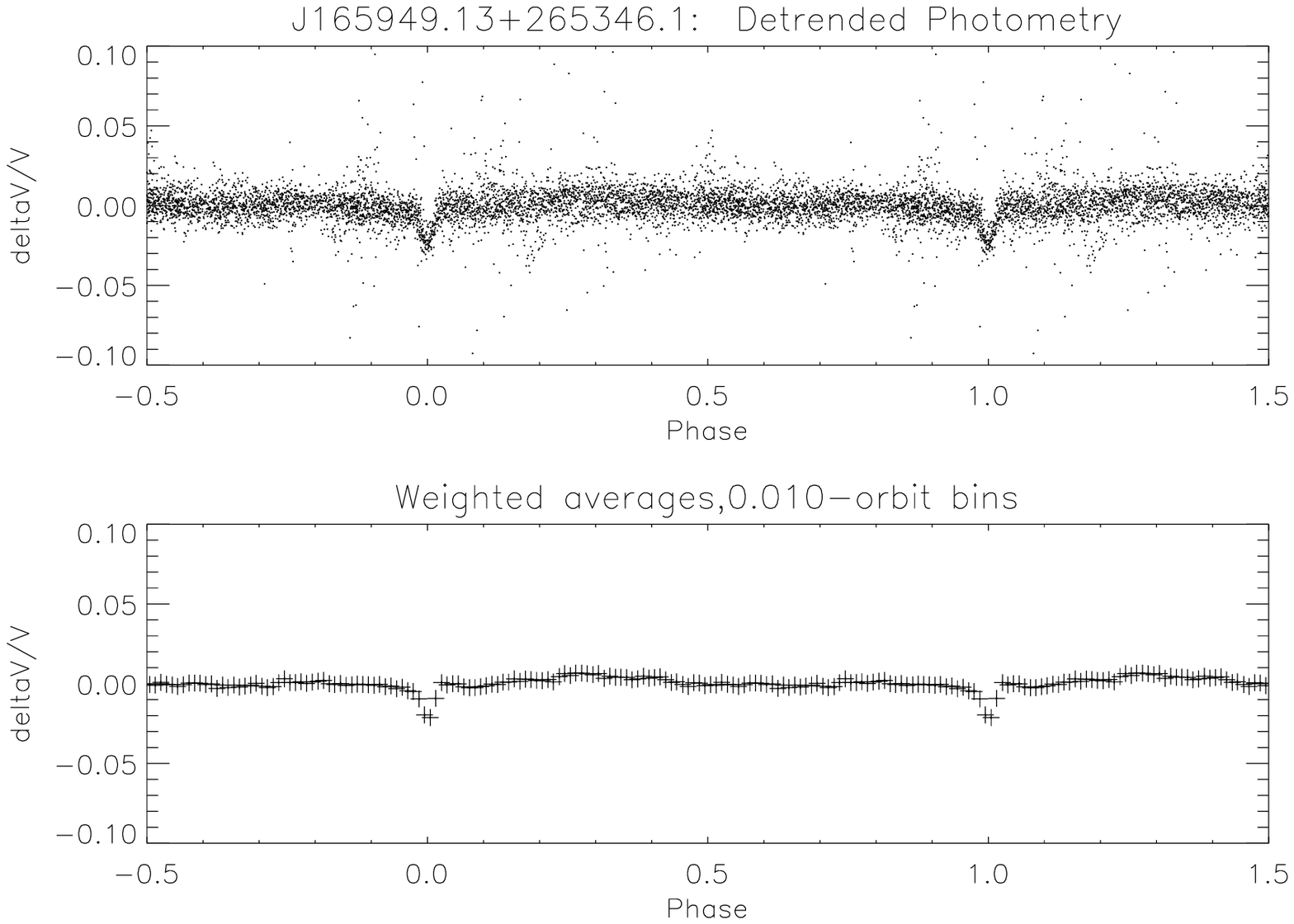} &
      \includegraphics[width=8.2cm]{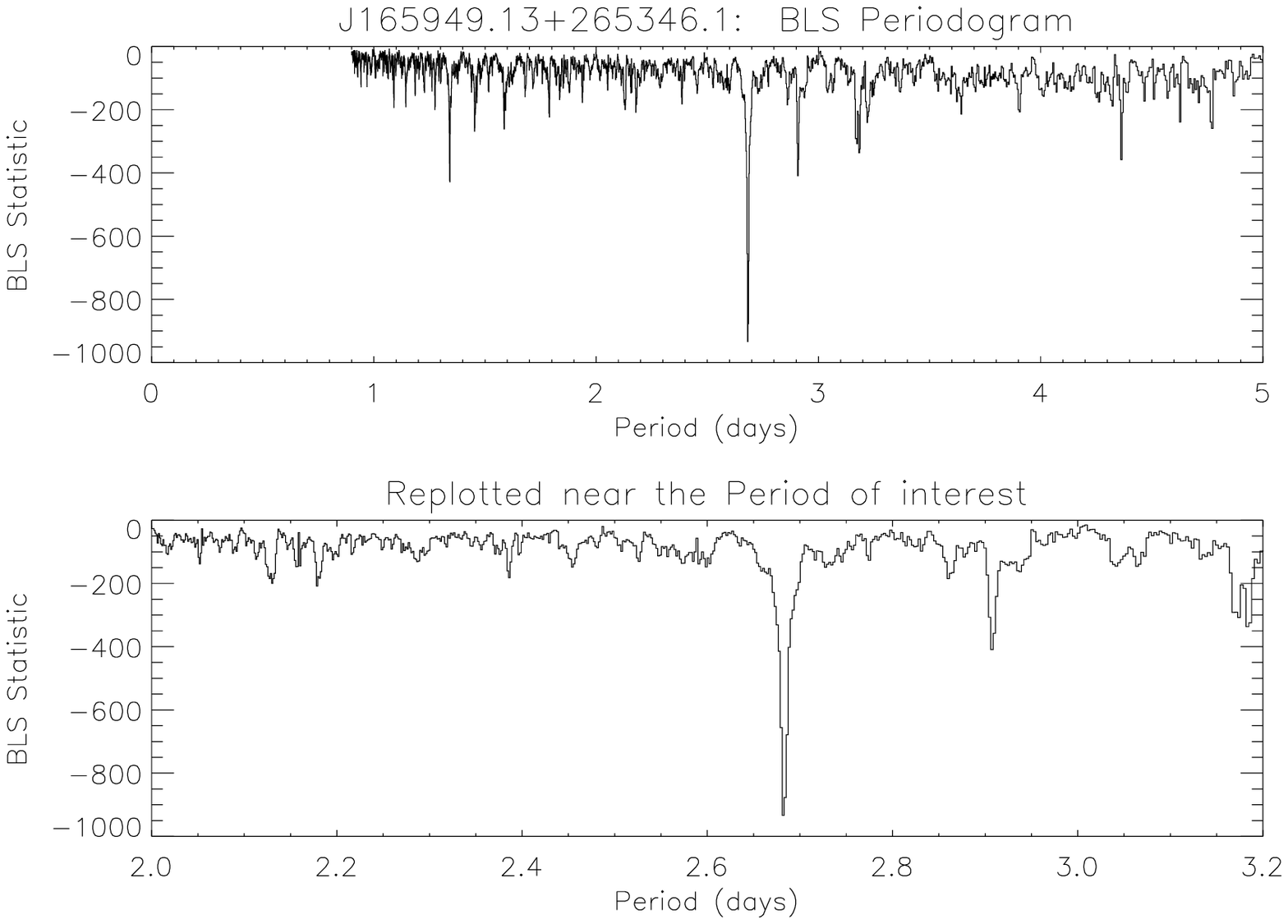} \\
    \end{tabular}
  \end{center}
  \caption{Example rejected candidates, showing the unbinned and binned
    lightcurves (left) and the BLS periodograms (right).}
\end{figure*}

{\bf 1SWASP J122557.80+334651.1:} This candidate was very high in the list
from the 12h fields, as can be seen from the high $\Delta \chi^2$ in Table
2. A total of 13 transits were observed and folding the data on the
strongest periodogram peak $\sim 1.36$ days yields a flat-bottomed transit
lightcurve, as shown in Figure 5. However, binning the data reveals the
out-of-eclipse variation which is further supported by the high value of
$\mathrm{S/N}_{ellip}$, although the value of $\Delta \chi^2 / \Delta
\chi_{-}^2$ is very high due to the depth of the eclipse. This candidate
was rejected on the basis of its high ellipsoidal variation.

{\bf 1SWASP J152645.62+310204.3:} A total of 16 transits observed with a
well-constrained period of $\sim 1.41$ days favoured this candidate. A
relatively high ellipsoidal variation, however, excluded this candidate
from passing through the second stage of selection criteria. Indeed, the
out-of-eclipse variation becomes especially apparent when the data is
binned.

{\bf 1SWASP J153135.51+305957.1:} This appears to be a promising candidate
with a strong signal to red noise ratio and 10 transits observed. The
flat-bottomed transits in the lightcurve folded on a period of $\sim 4.47$
days appear to be very convincing and the candidate passed through to the
third stage of candidate sifting. Even though the exoplanetary disgnostic
$\eta_p$ is unity for this candidate, the estimated size of the planet
based upon the spectral type is 2.03 $R_J$ and so was excluded from
appearing in the final list. A brown dwarf companion, though possible, is
only likely for relatively young ages where ongoing contraction allows for
a large radius \citep*{sta06}. Detection of low-mass stellar companions
such as OGLE-TR-122b \citep{pon05} and HAT-TR-205-103 \citep{bea07} have
shown that even low mass stars can have radii comparable to or even less
than giant planets. This target will be the subject of further
observations in a low-mass eclipsing binary study.

{\bf 1SWASP J160944.95+202609.7:} A large transit depth, or which 16 were
observed, contributed to this star having a very high $\Delta \chi^2$ and
$\Delta \chi^2 / \Delta \chi_{-}^2$. Additionally, the period of $\sim
1.64$ days is well determined by the strong peak in the periodogram.
However, the transits are distinctly ``V-shaped'' in appearance and the
ellipsoidal variation is slightly too large to prevent exclusion from the
list of candidates.

{\bf 1SWASP J163245.61+321754.9:} This candidate was one of the strongest
candidates selected by the detection algorithm and the subsequent visual
inspection. The $\Delta \chi^2$, $\Delta \chi^2 / \Delta \chi_{-}^2$, and
signal to red noise ratio are all exceptionally high. Indeed, the relative
strength of the primary peak in the periodogram at a period of $\sim 3.54$
days is striking. The transits are flat-bottomed and there is little
evidence of out-of-eclipse variation for the 10 eclipses observed, although
there is slight evidence for a secondary eclipse in the binned lightcurve.
The measured ellipsoidal variation for this star is more than enough for it
to be excluded from the candidate list, strengthening the case of this
candidate being an eclipsing binary star.

{\bf 1SWASP J165949.13+265346.1:} This lightcurve is a good example of a
transit mimic in that it is a subtle dip in the lightcurve with a small
depth, which is generally the signature one would expect from a transiting
planet. The 11 observed transits folded on the period of $\sim 2.68$ days
look convincing even when binned. However, once again the ellipsoidal
variation reveals that this star is also likely to be either a blend of an
eclipsing binary or a grazing eclipsing binary system.

\section{Discussion}

The preceding sections have described how 130,566 stars were extracted
from 729,335 for transit searching; and how the yield of 5,445 candidates
was reduced to a list of 6 candidates through the stringent selection
criteria. The criteria were largely designed to aggressively remove the
primary source of false-alarms, eclipsing binary stars, from the candidate
list. In this sense, the criteria proved to be very successful since, for
example, the ellipsoidal variation criteria dealt a devastating blow
against the kinds of false-alarms during the second stage and was the
major cause for elimination. Though this removed some very promising
looking candidates from the list, the evidence presented by closer
examination of a subset of the rejected candidates shows that there is
indeed clear eclipsing binary behaviour in the binned lightcurves, if not
in the unbinned data. The major source of elimination during the third
stage of candidate sifting was an excessive estimate of the planet size,
also generally due to an eclipsing binary star. Some flexibility was
allowed in the size criteria, particularly in view of the recent detection
of the large exoplanet TrES-4 \citep{man07}.

The detection of XO-1b by the transit detection algorithm and the
subsequent passing of all the selection criteria is an important test for
the transit sifting process. However, the recent discovery of HAT-P-3b
\citep{tor07} was cause for concern since it was observed in one of the
13h fields and has the identifier 1SWASP J134422.58+480143.2. The period
of $\sim 2.9$ days is within the parameter space which was searched by
the detection algorithm. Based upon the ephemeris information provided by
\citet{tor07}, the star was observed numerous times by SuperWASP-N during
at least 5 predicted transits during the 2004 observing season. However,
examination of the data show that the S/N for this star is exceptionally
low which resulted in a correspondingly low $\Delta \chi^2$ during the
fitting process. Combining the data with that from the 2006/2007 seasons
will undoubtedly yield a better result for this star.

Considering the large amount of sky surveyed in the RA range presented in
this paper, it is worth investigating the number of extra-solar planet
candidates one should expect and the practical limitations on achieving
this number. Analysis of radial velocity surveys such as \citet{san03}
and \citet{fis05} has shown that planet host-stars are preferentially
higher in metallicity. The field stars surveyed in these fields are
predominantly F--G--K dwarfs in the solar neighbourhood and so solar
metallicity is a reasonable approximation. Based upon the Besan\c{c}on
model \citep{rob03} constructed by \citet{smi06}, $\sim 46$\% of the
stars in typical SuperWASP fields are of F--G--K type and are therefore
of small enough size to produce detectable transit dips. Of the 130,566
stars in the 06h--16h fields searched for transits, around 60,000 stars
will meet this criteria.

It has been noted before by such papers as \citet{kan05} that the
frequency of hot Jupiters and the geometric consideration of randomly
oriented orbits will result in $\sim 0.1$\% of stars having an
observable transiting planet in a $1 < P < 5$ day orbit. This produces
an estimate of 60 transit candidates for the fields considered in this
paper. However, despite the large amount of sky coverage in the 06h--16h
RA range and the correspondingly large amount of stars monitored, the
largest constraint on these data is the lack of baseline coverage which
severely impedes the ability to detect multiple transits. The
probability plots shown in Figure 1 demonstrate this clearly with even
the 14h fields only having a 20\% chance of observing 4 transits of a
5 day period planet. Given this limitation for transit hunting in this
dataset, the final number of 6 transit candidates is not an
unreasonable detection rate. The eventual combination of several years
of SuperWASP data will greatly help to alleviate this deficiency for
this particular RA range.

Though the criteria used to remove eclipsing binary stars was generally
very successful, the problem of blended eclipsing binaries (as discussed
by \citet{bro03}) is more difficult to solve. \citet{odo06} gives a
particularly tricky example in which the light from a K dwarf binary
system was blended with the light from a late F dwarf star. Avoiding
confusion with these kinds of systems requires the use of careful
spectroscopic follow-up to identify blended light. The strategy adopted
for SuperWASP targets was to obtain high-resolution spectroscopic
snapshots of high priority candidates for fast and efficient elimination
of blends (described in detail by \citet{str07} and \citet{lis07}). The
follow-up campaign during 2006/2007 is continued photometric monitoring,
medium-resolution spectra for blend elimination, two-colour precision
photometry at the predicted times of transit, and finally precision
radial velocity monitoring using such instruments as SOPHIE.

\section{Conclusions}

This paper has described the acquisition, analysis, and results from the
transit-hunting SuperWASP-N 2004 observing campaign covering a right
ascension range of 06hr $<$ RA $<$ 16hr. Even though 56 fields were
monitored during this campaign, the lack of baseline coverage resulted in
the rejection of 23 fields from which 130,566 stars were extracted to be
searched for transit signatures. The selection criteria which were applied
to the resulting candidates proved to be exceptionally powerful at
removing eclipsing binaries from the candidate list due to the ellipsoidal
variations and the estimated planet size based upon the transit depth.
Amongst the final candidates is the known transiting planet XO-1b which
further demonstrates the strength of the transit sifting process. 

From the final list of 6 transit candidates, the photometry alone
indicates that 1SWASP J160242.43+290850.1 is in fact likely to be
the signature of a eclipsing binary rather than a planet. Furthermore,
follow-up spectroscopy of 1SWASP J161732.90+242119.0 shows that there
is no significant radial velocity variation, resulting in its rejection
as a planet candidate. Further follow-up observations, both photometric
and spectroscopic, are being undertaken for these and other fields from
the 2004 SuperWASP observing campaign. However, the real strength of
the 06--16h range will be realised when the 2004 data is combined with
that of subsequent years to create an exceptional baseline for transit
hunting.

\section*{Acknowledgements}

The WASP consortium consists of representatives from the Queen's
University Belfast, University of Cambridge (Wide Field Astronomy Unit),
Instituto de Astrofisica de Canarias, Isaac Newton Group of Telescopes
(La Palma), University of Keele, University of Leicester, Open
University, and the University of St Andrews. The SuperWASP-N instrument
was constructed and operated with funds made available from the
Consortium Universities and the Particle Physics and Astronomy Research
Council. SuperWASP-N is located in the Spanish Roque de Los Muchachos
Observatory on La Palma, Canary Islands which is operated by the
Instituto de Astrof\'isica de Canarias (IAC). The data reduction and
analysis described in this made extensive use of the Starlink Software
Collection, without which this project would not have been possible.
This research also made use of the SIMBAD database and VIZIER catalogue
service, operated at CDS, Strasbourg, France. In addition we made use of
data products from the Two Micron All Sky Survey, which is a joint project
of the University of Massachusetts and the Infrared Processing and
Analysis Center/California Institute of Technology, funded by the National
Aeronautics and Space Administration and the National Science Foundation.


\begin{thebibliography}{}

\bibitem[\protect\citeauthoryear{Beatty et al.}{2007}]{bea07} Beatty, T.G., et
  al., 2007, ApJ, 663, 573
\bibitem[\protect\citeauthoryear{Brown}{2003}]{bro03} Brown, T.M., 2003,
  ApJ, 593, L125
\bibitem[\protect\citeauthoryear{Burke et al.}{2006}]{bur06} Burke, C.J.,
  Gaudi, B.S., DePoy, D.L., Pogge, R.W., 2006, AJ, 132, 210
\bibitem[\protect\citeauthoryear{Burke et al.}{2007}]{bur07} Burke, C.J., et
  al., 2007, ApJ, in press
\bibitem[\protect\citeauthoryear{Cabanela et al.}{2003}]{cab03} Cabanela, J.,
  Humphreys, R., Aldering, G., Larsen, J., Odewahn, S., Thurmes, P., Cornuelle,
  C., 2003, PASP, 115, 837
\bibitem[\protect\citeauthoryear{Christian et al.}{2006}]{chr06} Christian D.J.,
  et al., 2006, MNRAS, 372, 1117
\bibitem[\protect\citeauthoryear{Clarkson et al.}{2007}]{cla07} Clarkson W.I.,
  et al., 2007, MNRAS, 381, 851
\bibitem[\protect\citeauthoryear{Collier Cameron et al.}{2006}]{col06} Collier
  Cameron, A., et al., 2006, MNRAS, 373, 799
\bibitem[\protect\citeauthoryear{Collier Cameron et al.}{2007a}]{col07a} Collier
  Cameron, A., et al., 2007a, MNRAS, 375, 951
\bibitem[\protect\citeauthoryear{Collier Cameron et al.}{2007b}]{col07b} Collier
  Cameron, A., et al., 2007b, MNRAS, 380, 1230
\bibitem[\protect\citeauthoryear{DeFa\"y, Deleuil, \& Barge}{DeFa\"y et
    al.}{2001}]{def01} DeFa\"y, C., Deleuil, M., Barge, P., 2001, A\&A, 365, 330
\bibitem[\protect\citeauthoryear{Fischer \& Valenti}{2005}]{fis05}
  Fischer, D.A., Valenti, J., 2005, ApJ, 622, 1102
\bibitem[\protect\citeauthoryear{Hartman et al.}{2004}]{har04} Hartman, J.D.,
  Bakos, G., Stanek, K.Z., Noyes, R.W., 2004, AJ, 128, 1761
\bibitem[\protect\citeauthoryear{H{\o}g et al.}{2000}]{hog00} H{\o}g, E.,
  et al., 2000, A\&A, 355, L27
\bibitem[\protect\citeauthoryear{Kane et al.}{2004}]{kan04} Kane, S.R.,
  Collier Cameron, A., Horne, K., James, D., Lister, T.A., Pollacco, D.L.,
  Street, R.A., Tsapras, Y., 2004, MNRAS, 353, 689
\bibitem[\protect\citeauthoryear{Kane et al.}{2005}]{kan05} Kane, S.R.,
  Collier Cameron, A., Horne, K., James, D., Lister, T.A., Pollacco, D.L.,
  Street, R.A., Tsapras, Y., 2005, MNRAS, 364, 1091
\bibitem[\protect\citeauthoryear{Lister et al.}{2007}]{lis07} Lister T.A.,
  et al., 2007, MNRAS, 379, 647
\bibitem[\protect\citeauthoryear{Konacki et al.}{2003}]{kon03} Konacki, M.,
  Torres, G., Jha, S., Sasselov, D., 2003, Nature, 421, 507
\bibitem[\protect\citeauthoryear{Kov\'acs, Zucker, \& Mazeh}{Kov\'acs et
    al.}{2002}]{kov02} Kov\'acs, G., Zucker, S., Mazeh, T., 2002, A\&A, 391, 369
\bibitem[\protect\citeauthoryear{Kov\'acs et al.}{2007}]{kov07} Kov\'acs, G.,
  et al., 2007, ApJ, 670, L41
\bibitem[\protect\citeauthoryear{Mandushev et al.}{2007}]{man07} Mandushev, G.,
  et al., 2007, ApJ, 667, L195
\bibitem[\protect\citeauthoryear{Monet et al.}{2003}]{mon03} Monet, D.G.,
  et al., 2003, ApJ, 125, 984
\bibitem[\protect\citeauthoryear{McCullough et al.}{2006}]{mcc06} McCullough,
  P.R., et al., 2006, ApJ, 648, 1228
\bibitem[\protect\citeauthoryear{Ochsenbein, Bauer, \& Marcout}{Ochsenbein
    et al.}{2000}]{och00} Ochsenbein, F., Bauer, P., Marcout, J., 2000, A\&AS,
  143, 23
\bibitem[\protect\citeauthoryear{O'Donovan et al.}{2006}]{odo06} O'Donovan,
  F.T., et al., 2006, ApJ, 644, 1237
\bibitem[\protect\citeauthoryear{O'Donovan et al.}{2007}]{odo07} O'Donovan,
  F.T., et al., 2007, ApJ, 663, L37
\bibitem[\protect\citeauthoryear{Pollacco et al.}{2006}]{pol06} Pollacco, D.L.,
  et al., 2006, PASP, 118, 1407
\bibitem[\protect\citeauthoryear{Pont et al.}{2005}]{pon05} Pont, F., Melo,
  C.H.F., Bouchy, F., Udry, S., Queloz, D., Mayor, M., Santos, N.C., 2005,
  A\&A, 433, L21
\bibitem[\protect\citeauthoryear{Pont, Zucker, \& Queloz}{Pont et
    al.}{2006}]{pon06} Pont, F., Zucker, S., Queloz, D., 2006, MNRAS, 373, 231
\bibitem[\protect\citeauthoryear{Protopapas, Jimenez, \& Alcock}{Protopapas et
    al.}{2005}]{pro05} Protopapas, P., Jimenez, R., Alcock, C., 2005, MNRAS, 362,
  460
\bibitem[\protect\citeauthoryear{Robin et al.}{2003}]{rob03} Robin, A.C.,
  Reyl\'e, C., Derri\`ere, S., Picaud, S., 2003, A\&A, 409, 523
\bibitem[\protect\citeauthoryear{Santos et al.}{2003}]{san03} Santos, N.C.,
  Israelian, G., Mayor, M., Rebolo, R., Udry, S., 2003, A\&A, 398, 363S
\bibitem[\protect\citeauthoryear{Sirko \& Paczy\'nski}{2003}]{sir03} Sirko, E.,
  Paczy\'nski, B., 2003, ApJ, 592, 1217
\bibitem[\protect\citeauthoryear{Skrutskie et al.}{2006}]{skr06} Skrutskie M.F.,
  et al., 2006, AJ, 131, 1163
\bibitem[\protect\citeauthoryear{Smith et al.}{2006}]{smi06} Smith A.M.S.,
  et al., 2006, MNRAS, 373, 1151
\bibitem[\protect\citeauthoryear{Stassun, Mathieu, \& Valenti}{Stassun
    et al.}{2006}]{sta06} Stassun, K.G., Mathieu, R.D., Valenti, J.A., 2006,
  Nature, 440, 311
\bibitem[\protect\citeauthoryear{Street et al.}{2007}]{str07} Street R.A.,
  et al., 2007, MNRAS, 379, 816
\bibitem[\protect\citeauthoryear{Tamuz, Mazeh, \& Zucker}{Tamuz et
    al.}{2005}]{tam05} Tamuz, O., Mazeh, T., Zucker, S., 2005, MNRAS, 356, 1466
\bibitem[\protect\citeauthoryear{Tingley \& Sackett}{2005}]{tin05}
  Tingley, B., Sackett, P.D., 2005, ApJ, 627, 1011
\bibitem[\protect\citeauthoryear{Torres et al.}{2004}]{tor04} Torres, G.,
  Konacki, M., Sasselov, D.D., Jha, S., 2004, ApJ, 614, 979
\bibitem[\protect\citeauthoryear{Torres et al.}{2007}]{tor07} Torres, G.,
  et al., 2007, ApJ, 666, L121
\bibitem[\protect\citeauthoryear{Wilson et al.}{2006}]{wil06} Wilson, D.M., et
  al., 2006, PASP, 118, 1245
\end{thebibliography}
\end{document}